\documentclass[twocolumn,twocolappendix%
    ,dvipsnames%,linenumbers%
    ]
    {aastex631}

\usepackage{graphicx}
\usepackage[utf8]{inputenc}

\usepackage{savesym}
\savesymbol{tablenum}  %.. avoid '\tablenum already defined' error
\usepackage{siunitx}
\restoresymbol{SIX}{tablenum}

\usepackage{mathrsfs, amsmath, amssymb, amsthm, amsxtra, bm}
\usepackage{multirow}
\usepackage[hyphens]{url}
\usepackage{hyperref}
\usepackage[nameinlink,capitalize,noabbrev]{cleveref}
\usepackage{todonotes}
\usepackage{svg}
\usepackage[section]{placeins}
\usepackage{pgfplotstable}
   \pgfplotsset{compat=1.18} 

\usepackage{colortbl}
\usepackage{xspace}
\AtBeginDocument{%
   \NewCommandCopy\autoreforig\autoref
   \RenewDocumentCommand{\autoref}{som}{%
      \IfBooleanF{#1}{%
         \hyperref[#3]%
      }{%
         \autoreforig*{#3}\IfValueT{#2}{\nobreak}\xspace\IfValueT{#2}{#2}%
      }%
   }
}

\usepackage{commands}
\usepackage{names}
\usepackage{seancmds}

\newcommand{\Csh}    {C_{\text{sh}}}

\newcommand{\init}   {0.17}%AU
\newcommand{\Epidot} {\dot{E}_\text{PI}}

\newcommand{\response}[1]{#1}

\DeclareSIUnit \parsec {pc}
\DeclareSIUnit \gauss {G}
\begin{document}

%\title{Sampling solar wind transport model parameters using MC analysis}
\title{Constraining solar wind transport model parameters using Bayesian analysis}

\author[0009-0002-8645-5139]{Mark A. Bishop}
\email{mark.bishop@vuw.ac.nz}
\affiliation{School of Chemical and Physical Sciences, 
Victoria University of Wellington,
Wellington 6012,
New Zealand}
\affiliation{Department of Mathematics,
University of Waikato,
Hamilton 3240,
New Zealand}

\author[0000-0002-2814-7288]{Sean Oughton}
\affiliation{Department of Mathematics,
University of Waikato,
Hamilton 3240,
New Zealand}

\author[0000-0003-0602-8381]{Tulasi N. Parashar}
\affiliation{School of Chemical and Physical Sciences, 
Victoria University of Wellington,
Wellington 6012,
New Zealand}

\author[0000-0002-6255-8240]{Yvette C. Perrott}
\affiliation{School of Chemical and Physical Sciences, 
Victoria University of Wellington,
Wellington 6012,
New Zealand}

\begin{abstract}
%\blue{recap the methodology.}\\
We apply nested-sampling Bayesian analysis
to a model
for the transport of magnetohydrodynamic-scale solar wind fluctuations.  
The dual objectives are to obtain improved constraints on  parameters present in the turbulence transport model (TTM) and to support 
quantitative comparisons of the quality of distinct versions of the transport model.
The TTMs analysed are essentially the one-dimensional steady-state ones presented in \cite{BreechEA08} that describe the radial evolution of 
the energy, correlation length, and normalized cross helicity of the fluctuations, 
together with the proton temperature, in prescribed background solar wind fields.
 %% from 
 %%  $ 0.17\,\si{AU}$ to $\sim 100\,\si{AU}$. 
Modelled effects present in the TTM include nonlinear turbulence interactions, shear driving, and energy injection associated with pickup-ions. 
Each of these modelled effects involves adjustable parameters that we seek to constrain using Bayesian analysis. We find that, given the TTMs and observational datasets analysed, the most appropriate TTM to recommend corresponds to two-dimensional fluctuations and has
von K\'arm\'an--Howarth parameters of 
 %\red{If we're going to cite the uncertainties (which I'm not keen on for the abstract) I think the $\approx$ should be replaced with =: \\}
 % $\alpha \approx 0.16$\red{$\pm 0.03$} and $\beta \approx 0.10$\red{$\pm 0.02$}, 
 $\alpha \approx 0.16$ and $\beta \approx 0.10$, 
along with reasonably standard values for the other adjustable parameters. 
The analysis also indicates that it is advantageous to include pickup ion effects in the lengthscale evolution equation by assuming 
 $ Z^{2\beta/\alpha} \lambda $ 
is locally conserved.
Such Bayesian analysis is readily extended to more sophisticated solar wind models, space weather models, and might lead to improved predictions of, for example, solar flare and CME interactions with the earth. 

\end{abstract}


    \section{Introduction}
      \label{sec:intro}
%\bigskip

The evolution of 
  magnetohydrodynamic (MHD) scale fluctuations
  %% fluctuations at magnetohydrodynamic (MHD) scales 
in the solar wind is a  topic of long-standing interest, with relevance to various aspects of space physics including 
space weather \citep{Schwenn06}, 
scattering of cosmic rays \citep{PeiEA10, Engelbrecht.etal22},
and their roles in connection with the global evolution of the heliospheric fields
  \citep[\eg{}][]{vanderHolstEA14, Pogorelov.etal24}.
Under various assumptions and approximations this evolution has been investigated observationally, theoretically, and numerically
  \citep[for recent reviews see][]{BrunoCarbone13, 
  OughtonEngelbrecht21, FraternaleEA22-ssr,
  SmithVasquez24}.
%.. PinEA24-iv
Early models were based on the WKB approximation and treated the fluctuations as MHD waves
    \citep{Parker65-wkb}.
Transport models that accounted for turbulence fluctuations (as well as waves) were presented in the 1980s 
  \citep{TuEA84,Tu87,VelliEA89,ZhouMatt90a,TuMarsch95}.
Extensions of these turbulence transport models (TTMs) 
support time-dependence, three-dimensionality, and
driving by stream shear and pickup ion induced waves, and interactions in the heliosheath and local interstellar medium.   
 %%  \blue{maybe include refs??? There are a lot!  And we give plenty of them later in the intro}

%   \citep{MattEA94-ec,MattEA96-jpp, MattEA99-swh, ZankEA96, SmithEA01,SmithEA06-pi, IsenbergEA03, BreechEA08,    UsmanovEA11, ZankEA12-xport,  ZankEA18-pui, UsmanovEA16, KleimannEA23}.

Many of these TTMs belong to the ``energy-containing'' class 
  \citep{MattEA94-ec,MattEA96-jpp, ZankEA96},
meaning that they evolve \emph{bulk} characteristics of the turbulence (such as its energy and correlation length), rather than tracking, for example, energy spectra.
These typically involve modelling of the nonlinear terms associated with turbulence and also of driving processes, usually with adjustable parameters being features of the modelling.  The question of how to best determine these adjustable parameters is a primary focus of this work.

Beyond 1\si{AU} or so,
agreement between observations and TTM simulation results is, typically, encouraging
  \cite[e.g.,][]{SmithEA01, BreechEA08, AdhikariEA21-xport, FraternaleEA22-ssr,SmithVasquez24}.  
However, the
  %%  adjustable
modelling parameters used in such studies have usually been estimated in rather \emph{ad hoc} fashion with few systematic efforts made to find optimal or best-fit values. 
For example, two common approaches are (i)
 generating `best-fitting' values based on guess-and-check, and then by-eye validation against solar wind observations, 
and (ii)
displaying several solutions corresponding to different values of the adjustable parameters
  \citep[\eg{}][]{ZankEA96, MattEA99-swh, SmithEA01, BreechEA08, OughtonEA11, UsmanovEA18, AdhikariEA15, AdhikariEA21-xport}. 
Although these approaches have had considerable success, it would of course be preferable to employ a systematic method for constraining any adjustable parameters present in the TTMs.  
Ideally this constraining method would be based on both the TTM and the solar wind data, and also provide a statistically rigorous way of comparing TTMs to one another. 
Bayesian analysis methodologies, 
 which determine (posterior) probability distributions for model parameters from imposed prior distributions and data,
can address both of these aspects.  
Our objective herein is to show how Bayesian analysis can be used to constrain and compare solar wind transport models.

The most sophisticated energy-containing TTMs are time-dependent, three-dimensional, and can provide (numerical)  solutions throughout the heliosphere and, in some cases, into the local interstellar medium
  \citep[e.g.,][]{UsmanovEA11, UsmanovEA14, UsmanovEA16, WiengartenEA15, WiengartenEA16, ShiotaEA17, KleimannEA23}.
See \cite{FraternaleEA22-ssr} for a review.
In many applications, however, the simplified situation of a steady-state, spherically symmetric, and one-dimensional (radial) heliosphere 
is of relevance 
  \citep[e.g.,][]{MattEA94-ec,ZankEA96,MattEA96-jpp,MattEA99-swh,SmithEA01,BreechEA08, OughtonEA11,
  AdhikariEA15, AdhikariEA17-NI-2, AdhikariEA21-xport, AdhikariEA23-highBetaXport}.
Naturally this affords considerable numerical advantages and even allows determination of some analytic solutions
 \citep{MattEA96-jpp, ZankEA96, OughtonBishop-sw16}.
Hereafter, 
we focus on this simplified steady 1D situation, applying Bayesian analysis to essentially the TTMs presented in \cite{BreechEA08}. 
 \response{Use of a simpler TTM facilitates
 a clean presentation of the approach, which to the best of our knowledge represents the first time Bayesian analysis has been used with a solar wind TTM.}

Bayesian methodologies have been used extensively in the fields of astronomy and cosmology \citep[see][for a review]{trotta2008bayes}. 
Well known examples include
constraining the Hubble constant on Type Ia supernovae data \citep{Perlmutter.etal95, Abbott.etal19}, and 
constraining the other cosmological constants based on cosmic microwave background observations \citep{Jungman.etal96, PlanckCollaboration.etal20}. 
On the astronomy side, applications have included
various profile fittings \citep{Perrott.etal19, Javid.etal19} 
and determination of many different astrophysical parameters
 \citep[see, e.g.,][and references therein]{Parkinson.Liddle13,Sharma17}.
  %% and determining parameters associated with pulsars \citep{Riley.etal19, Riley.etal21}, exoplanet transits \citep{Espinoza.etal19}, exoplanet atmospheres \citep{Tsiaras.etal19, Line.etal21}, cosmic neutrino detection \citep{IceCubeCollaboration.etal21}, gravitational wave detection \citep{Veitch.etal15, Williams.etal21, Williams.etal23}, and gravitational wave background detection \citep{Agazie.etal23}.  For reviews see \cite{Parkinson.Liddle13} and \cite{Sharma17} and references therein. 

 %   \red{... (sean) 1. Have commented out most of the specific example references in above para.  But if you think they should be there, uncommenting will reinstate them.\\
 %   2. Do we need the next sentence (sean)? \\
 %  There has been a lot of work and understanding in the field of cosmology on Bayesian methodologies, due to the current $H_0$ \citep{Verde.etal19, Wong.etal20, Freedman21} and $S_8$ tensions \citep{Poulin.etal23, Joseph.etal23}.
 % }

  %% Bayesian analysis 
These techniques have also been used in assorted other disciplines. 
See \cite{Ashton.etal22} for a review with a focus on the nested sampling method that we employ herein
  (cf.~\autoref{appendix:mcmc})
and \cite{Ellison04}, \cite{vontoussaint11}, \cite{Huang.etal19}, and \cite{vandeschoot.etal21} 
for more general Bayesian inference application reviews. 

As a consequence of this extensive use of Bayesian analysis, many software packages have been developed, including ones designed explicitly for applications in astronomy and astrophysics \citep{Feroz.etal09, Brewer.etal10, Corsaro.DeRidder14, Speagle20, Barbary21, Kester.Mueller21, Buchner21} 
and cosmology \citep{Hojjati.etal11, Parkinson.etal11, Handley.etal15}.
In performing our analysis we will make use of some of this public domain software.

  %% Bayesian methods have been used to help assess the role of turbulence in astronomical observations. \cite{vogt_measuring_2003, vogt_bayesian_2005, bonafede_2010_coma} use Bayesian methods to fit a fractional Brownian motion field to Faraday rotation measure observations, giving a constraint on the fluctuations in the magnetic fields of the intracluster medium \citep[see also,][]{govoni_intracluster_2006, guidetti_intracluster_2008, vacca_intracluster_2010, vacca_intracluster_2012}. Bayesian methods have also been used for fitting powerlaw slopes to an observed power spectrum of X-ray surface brightness fluctuations \citep{eckert_deep_2017, mirakhor_deep_2023}.

%% \response{One objective of our work here is to demonstrate the application of Bayesian analysis to a TTM, and specifically to one of the simpler ones to support a clean presentation.  Its application to more complete and/or sophisticated TTMs will be considered in future work.}

The structure of the paper is as follows.  In \autoref{sec:TTM} we summarize the TTMs and observational datasets to which we apply Bayesian analysis.  The Bayesian analysis technique is outlined in \autoref{sec:bayes_analysis} with the results of the analyses presented in \autoref{sec:results}. 
\autoref{sec:discussion} contains discussion of the results, followed by a conclusion section.  
Two appendices close the paper, presenting further details regarding 
the TTMs (\autoref{appendix:breech_model}) 
and Bayesian methodologies (\autoref{appendix:mcmc}).

 %% Bayesian techniques have also been applied to other disciplines such as: 
 %% systems biology \citep{Wilkinson07, Lillacci.Khammash10, Pullen.Morris14, Reali.etal17, Valderrama-Bahamondez.Frohlich19, Mikelson.Khammash20}, 
 %% phylogenetics \citep{Yang.Rannala97, Russel.etal19}, 
 %% materials science \citep{NS_pymatnest_2017, NS_review}, and 
 %% signal processing \citep{Fitzgerald01, Doucet.Wang05, Henderson.etal17}.

\section{The Turbulence Transport Models}
\label{sec:TTM}

\subsection{Model Equations}
    \label{sec-TTMeqns}

We employ a previously developed 
  \citep{BreechEA08}
%% turbulent transport model (TTM) 
 TTM
that assumes a one-dimensional (radial $r$) 
%% neglecting the $\theta$ and $\phi$ components), 
steady-state large-scale solar wind 
with  uniform flow speed $ U \hat{\vr}$,
density $ \rho \propto 1/r^2$,
and Alfv\'en speed $ \vV_A(r) $. 
Three quantities characterising the incompressible turbulence fluctuations are followed---energy $Z^2$, 
correlation length $\lambda$, 
and normalized cross helicity $\sigma_c$---plus 
the proton temperature $T$. 
%% Assuming the the ratio of \alfven{} speed to solar wind speed (\ie{}terms of $\mathcal{O}(V_A/U)$) is small, 
We solve the TTM in the super-Alfv\'enic flow region, 
neglecting terms of order $V_A/U$ to obtain
(cf.\ \autoref{appendix:breech_model}):
\begin{align}
  \label{eq:model-Z2}
   U \Deriv{Z^2}{r}  &= 
    - \frac{UZ^2}{r} \left[
           1 +  \sigma_D M_Z - \Csh
       \right]
    + \Epidot
    - \alpha f^+(\sigma_c) \frac{Z^3}{\lambda } ,
 \\
  \label{eq:model-lambda}
  U \Deriv{ \lambda} {r}  &=
         \frac{\lambda \sigma_D M_\lambda U} {r} 
      +  \beta f^+(\sigma_c) Z
      -  \frac{\beta}{\alpha} \frac{\lambda}{Z^2} \Epidot,
\\
  \label{eq:model-sigc}
   U \Deriv{\sigma_c} {r}  &= 
     \alpha f'(\sigma_c) \frac{Z }{ \lambda}
     -\left[\frac{U}{r} \left( \Csh - \sigma_D M_Z \right)
        +  \frac{\Epidot} {Z^2}
      \right] \sigma_c ,
 \\
  \label{eq:model-temp}
   U \Deriv{T} {r} &= 
      -\frac{4}{3}\frac{TU}{r}
      + \frac{\alpha}{3}\frac{m_p}{k_B}
         f^+(\sigma_c) \frac{Z^3}{\lambda} ,
\end{align}
where 
$ f^\pm(\sigma_c) = \sqrt{1 - \sigma_c^2} 
 \left[ \sqrt{1 + \sigma_c} \pm \sqrt{1 - \sigma_c} \right] /2 $, 
$f'(\sigma_c) = \sigma_c f^+(\sigma_c) - f^-(\sigma_c)$,
%\begin{align}
%    f^\pm(\sigma_c) & = \frac{\sqrt{1 - \sigma_c^2}}{2} \bracket{\sqrt{1 + \sigma_c} \pm \sqrt{1 - %\sigma_c}},
%  \label{eq:fpm}
%  \\
%   f'(\sigma_c) &= \sigma_c f^+(\sigma_c) - f^-(\sigma_c)
%    .
%  \label{eq:fprime}
%\end{align}
$m_p$ is the proton mass, $k_B$ is Boltzmann's constant, and the other quantities are discussed below.
Terms in the equations represent physics associated with advection by the mean wind, large-scale gradient `mixing' effects, and modelling of energy dissipation and driving
  \citep{BreechEA08}.
The TTM is appropriate for any helio-colatitude $\theta$, although here 
  %% herein 
we restrict our analysis to ecliptic cases, i.e., $\theta = \pi/2$.

One sees that the TTM depends on the so-called `mixing operators', 
    $ M_Z $ and $ M_\lambda $,
that take different forms depending upon the underlying symmetry of the fluctuations. 
 We suppose that the fluctuations have one of two distinct symmetries:
either 3D isotropic or  2D isotropic, 
where the latter means the fluctuation amplitudes are restricted to be perpendicular to the mean 
   (Parker spiral)
magnetic field and to be isotropically distributed 
in those 2D planes.
For brevity we refer to these two cases as the 3D and 2D models.
Under the stated assumptions one obtains
  \citep{MattEA94-ec,MattEA96-jpp, ZankEA96, BreechEA08}, 
\begin{align}
    \label{eq:Mops_2D}
   \text{3D model: } \quad & M_Z = \frac{1}{3}, &  M_\lambda &= \frac{1}{3},
  \\
   \text{2D model: } \quad & M_Z = \cos^2 \psi, &  M_\lambda &= \sin^2 \psi 
    ,
 \label{eq:Mops}
\end{align}
where $\psi(r)$ is the angle between the mean interplanetary magnetic field, taken to be a \cite{Parker58} spiral,
and the radial direction.
 % \begin{align}
 %   \psi &=  \arccos\bracket{\frac{V_{r}}{\sqrt{V_{r}^2 + V_{\phi}^2}}} , %- \arccos\bracket{\unitvec{V}_{A} \cdot \unitvec{r}} .%= - \arccos\bracket{\frac{V_{r}}{\sqrt{V_{r}^2 + V_{\phi}^2}}} ,
 % \end{align}
 % with $V_r$ and $V_\phi$ the radial and azimuthal components of the large-scale Alfv\'en velocity, 
 %    $ \vV_A = \vB_0 / \sqrt{4\pi\rho} $.
 %% \red{The negative sign is to ensure that the angle applies a rotation \emph{clockwise} from a magnetic field aligned coordinate system, to a radially aligned coordinate system.}
For definiteness we employ
the Parker spiral associated with an Alfv\'en radius 
  %% $r_A$
 %% (the distance where the mean solar wind speed becomes equal to the mean Alfv\'en speed) 
of 10 solar radii, and a solar rotation rate of $\qty{2.9e-6}{rad/s}$. 

Also present in the TTM are five adjustable parameters: 
  $\alpha$, $\beta$, $\Csh$, $f_D$, $\sigma_D$.
It is these, 
along with the four  inner  boundary conditions (BCs), 
  %%   at $ r= \qty{0.17}{AU} $,
that we will use Bayesian analysis to constrain.
Physically the parameters relate to 
  von K\'arm\'an--Howarth (vKH) style 
modelling of the turbulent energy cascade ($\alpha$ and $\beta$),
modelling of stream shear driving ($\Csh$) and pickup ion driving 
  ($f_D$,  see \autoref{eq:Epi-dot}),
and
a closure for the energy difference 
  ($\sigma_D$).
  %% , see next paragraph).
    %% \autoref{eq:dsigDdr}).
In the next subsection we discuss several points concerning these parameters and the application of Bayesian analysis to them.
For further discussion of the physics associated with the modelling and adjustable parameters see, for example,
\cite{BreechEA08}.

\response{To provide a clean presentation of the Bayesian analysis approach for determining the TTM parameters we have chosen to use relatively simple TTMs.  
(Application to more complete and/or sophisticated TTMs, as referenced in \autoref{sec:intro}, is left for future work.)
For example, we employ a uniform wind speed of 
$ U = \qty{400}{km/s}$ 
and neglect features such as 
transport terms of order $V_A / U$ 
and the decrease of $U(r)$ with distance
  \citep{WangRichardson03, IsenbergEA10, ZankEA18-pui, ElliottEA19}.
Furthermore, 
the modelling of energy injection, associated with either velocity shear (at constant correlation length) or pickup ion instabilities,
is rather simple, with more developed modelling approaches available
  \citep[e.g.,][]{ZankEA12-xport, WiengartenEA15, Isenberg05}.
  Such extensions can be readily incorporated.
We remark that the modelling approaches for the shear and pickup ion driving differ; see 
  \cite{BreechEA08} for more details.
}

\response{We have also 
made the simplifying assumption that all of the cascaded turbulence energy is used to heat solar wind protons.  Of course this is unlikely to be realistic, with studies indicating that about 40\% of the turbulence energy heats solar wind electrons instead \citep{BreechEA09, CranmerEA09, Howes11, BandyopadhyayEA23-peHeat}. Relaxation of this `proton only' heating assumption is considered in \autoref{appendix:aT_analysis}.}

%---------------------------------------------------

    \subsection{Estimation of Adjustable Parameters}
    \label{sec:adjustables}

As noted in the Introduction,  
  %% \autoref{sec:intro}
previous studies have often estimated adjustable parameters in a TTM in somewhat \emph{ad hoc} or non-systematic ways.
Here we give brief descriptions of some of these methods and indicate how one may instead apply Bayesian analysis.

%\green{ Typically, a von \karman{}--Howarth (vKH) phenomenology \citep{Taylor35, KarmanHowarth38, Dryden43} has been used as closure for the nonlinear term of the Navier-Stokes, second order correlation function equations. Similarly, applied to the \elsasser{} form of the magnetohydrodynamic equations \citep{ZhouMatt90a, MattEA94-ec}; describing the rate of change from the nonlinear terms for the (\elsasser{}) energy and correlation lengths ($Z^2$ and $\lambda$):
% \begin{align}
%     \Deriv{Z^2} {t}\bigg|_{NL} &= 
%     - \alpha \frac{Z^3}{\lambda} ,
%     \label{eq:dZ2dt}
%  \\
%     \Deriv{\lambda} {t}\bigg|_{NL} 
%      &= 
%      \beta Z ,
%   \label{eq:dldt}
% \end{align}
% where $\alpha,\,\/\beta \ge 0$ are left as (MHD) `vKH parameters' that play an important role in the energy cascade process \citep{MattEA94-ec, MattEA96-jpp, WanEA12-vKH}.  
% Various theories of turbulence evolution can be (re)constructed via specific choices of $\alpha$ and $\beta$ 
% \citep{MattEA96-jpp}, 
% including, for example, Kolmogorov's model \citep{Kol41b} with $\alpha = 5 \beta$.}

 %% comparing the behaviour of the exact energy decay (in homogeneous MHD) with that for the vKH decay phenomenology:
 %     $ \d Z^2 / \d{t} = - \alpha Z^3 / \lambda $
 %   \cite{HossainEA95,LinkmannEA17-Ceps,BandyopadhyayEA18-Ceps}.
 
  %% Alternatively, $\alpha$ and $\beta$ may be chosen based on analysis of simulation results \citep{HossainEA95, LinkmannEA17-Ceps, BandyopadhyayEA18-Ceps, BandyopadhyayEA19-lengths}. 

Let us start with 
  the vKH parameters 
$ \alpha$ and $\beta$.
Values for these have sometimes been chosen on the basis of results from homogeneous decaying MHD turbulence simulations
\citep{HossainEA95, LinkmannEA17-Ceps, BandyopadhyayEA18-Ceps, BandyopadhyayEA19-lengths}.
In other cases
$\alpha$ has been set by extending the equivalent experimentally-determined quantity for homogeneous decaying Navier--Stokes turbulence \citep{Sreenivasan98} to MHD 
   \citep{UsmanovEA14,WrenchEA24}.
   %.. App B and A, respectively.
In either situation, a value for $\beta$ can be assigned by imposing one of the conservation laws from the family applicable in homogeneous turbulence cases, 
  $ Z^{2\beta/\alpha} \lambda = \mathrm{const} $,
with the ratios $2\beta / \alpha = 1$ or 2 of particular physical significance
   \citep{HossainEA95, MattEA96-jpp, WanEA12-vKH, ZankEA96,ZankEA12-xport}.
These are of course physically well-motivated approaches. However, they are unlikely to provide values that can provide optimal fits to data,
motivating our application of Bayesian analysis to determine posterior probability distributions for these parameters.
 
  %% \blue{The vKH parameters provide a good starting point and initial motivation for applying the Bayesian analysis.}

Consider now
the $\sigma_D$ parameter appearing in equations~\eqref{eq:model-Z2}--\eqref{eq:model-sigc}. 
This arises because,
in place of a dynamical equation for the energy difference  
 $D = E^v - E^b $ (aka residual energy), 
we employ the approximation that the normalized energy difference, 
  $ \sigma_D = D / Z^2 
    \equiv (E^v - E^b) / (E^v + E^b) 
  $, 
is a constant
  ($ E^v, E^b $ are the fluctuation kinetic and magnetic energies).
On the basis of 
solar wind observations \citep[e.g.,][]{PerriBalogh10-sigc}  the choice $\sigma_{D} \approx -1/3$ has frequently been used in TTMs \citep[e.g.,][]{ZankEA96, MattEA96-jpp, SmithEA01, BreechEA08, YokoiEA08, OughtonEA11}. 
In \autoref{sec:extended_analysis} we will employ Bayesian analysis to investigate the appropriateness of using
    $ \sigma_D = \mathrm{const} $.
Moreover, even though we treat it as a constant, we shall see 
in \autoref{sec:bayes_analysis}
that it is advantageous to nonetheless include $\sigma_D$ in the Bayesian analysis likelihood calculations 
 (see \autoref{sec:bayes_analysis}). 
    %% This enables comparison of the TTM's \emph{constant} value of $\sigma_D$ with the (varying) observational $\sigma_D$ data. 
Implementation-wise, this is equivalent to adding the  evolution equation 
\begin{align}
    \Deriv{\sigma_D} {r} = 0 
  \label{eq:dsigDdr}  
\end{align}
to the system Eqns~\eqref{eq:model-Z2}--\eqref{eq:model-temp};
clearly this leads to $\sigma_D$ being equal to its inner boundary value. 
  %%    $\sigma_{D,\init} $. 
%  \red{ \\ NEXT SENTENCE TOO DETAILED FOR HERE (so just drop it)?: \\}
%In our Bayesian analyses which do not produce a posterior distribution for $\sigma_D$ 
%(e.g., \autoref{sec:ab_analysis})
%we employ the observationally motivated value 
%  $ \sigma_D = -1/3 $ 
%  (\autoref{tab:common_model_parameters}).

%\red{Really, we should also be constraining $\sigma_D$ based on the Adhikari data. This will mean I will need to add a constant equation for $\sigma_D$ that just compares its constant value to the data for the Bayesian likelihood calculation. This will mean we can compare the evidences for constraining a constant $\sigma_D$ versus not.} \textcolor{olive}{This takes a bit more work, as I treat this as a evolution equation, with zero derivative, which enables to log-likelihood calculation based on $\sigma_D$ data, this unfortunately means I will need to rerun ALL of the analysis because the log-evidences won't be comparable.} \textcolor{teal}{I have run the $\alpha$-$\beta$-$T_{\init}$ and $\alpha$-$\beta$-all analysis with the \emph{constant} $\sigma_{D}$ (really, $\sigma_{D,0}$, with zero derivative) and found that the mean value is consistently $\sigma_{D,0} \approx -0.37$, which is satisfyingly close to $\sigma_{D,0} = -1/3$ from literature. I think this is a worthwhile result to publish too.} \textcolor{blue}{Tulasi made a wise comment, that I should leave this analysis for another time.}

Estimates for $\Csh$, 
the constant determining the strength of the driving due to large-scale velocity shear
  \citep{Burlaga74, SmithWolfe76, NeugebauerEA95},
have primarily been set using simple theoretical models of shear evaluated using typical observational values
  \citep[e.g.,][]{ZankEA96,MattEA99-swh,SmithEA01,BreechEA08}.  
There is thus some similarity with the approach used in determining $\sigma_D$. 
\response{We follow \cite{BreechEA08} in modelling the shear interaction between faster and slower streams of the solar wind. In particular, the shear driving is assumed to occur at the correlation length scale and thus there is no explicit shear driving term in the $\lambda$ equation~\eqref{eq:model-lambda} \citep{BreechEA05}.}

The final adjustable parameter to consider, $f_D$, 
controls the overall strength of the energy injection associated with waves generated by the scattering of pickup ions.
Here we model the process using 
  \citep{WilliamsZank94, ZankEA96, MattEA99-swh, SmithEA01, BreechEA08}:\footnote{Note that  
    \autoref{eq:Epi-dot}
    is an approximate model and more rigorous expressions have been determined and employed in TTMs
    \citep{IsenbergEA03, Isenberg05, SmithEA06-pi, OughtonEA11}.}
\begin{align}
    \frac{\Epidot} {U} = f_D \frac{V_A n^\infty_H}{n^{1 \si{AU}}_{SW} \tau_{\text{ion}}^{1\si{AU}}} 
    {\rm e}^{-L_{\text{cav}}/r} ,
  \label{eq:Epi-dot}
\end{align}
where 
$n^\infty_H$ is the interstellar neutral hydrogen density, 
$n_{SW}^{1\si{AU}}$ is the solar wind density at $1\,\si{AU}$, 
$L_{\text{cav}}$ defines the ionization cavity within which minimal pickup ion effects occur, and 
$\tau_{\text{ion}}^{1\si{AU}}$ is the ionization timescale at $1\,\si{AU}$.
These quantities are reasonably well determined and we use commonly chosen values for them
   (see \autoref{tab:common_model_parameters} 
     and, e.g., \citealt{BreechEA08}).
Studies that 
    employ
 %%  make use of
\autoref{eq:Epi-dot}
have often explored values of $ f_D \sim 1/10 $ but usually without efforts to determine an optimum $f_D$ 
    \citep{ZankEA96,MattEA99-swh,SmithEA01,BreechEA08,UsmanovEA11}.
Hence, this parameter is also a suitable candidate for investigating with Bayesian analysis.

   %% The PI terms account for the additional driving of turbulence due to the generation of waves associated with the ionization of neutral interstellar hydrogen in the outer heliosphere via photoionization, or charge exchange. 

 %  \red{We have chosen $n_{SW} = 5$\,cm$^{-3}$ which is actually the value of $n_{SW}$ at \qty{1}{AU}, NOT \qty{0.3}{AU}}
 %  \textcolor{teal}{This will probably have minimal effect}.

% Turning to the shear driving,
% although we treat the solar wind speed as constant everywhere, as is well known it actually varies with helio-latitude (and distance) allowing for shear interactions between faster and slower streams. This interaction creates instabilities which can add energy to the system.
% We follow \cite{BreechEA08} in modelling this.
% The $\Csh$  parameter represents the strength of the shearing interaction which depends on the difference of fast-to-slow solar wind speeds ($\Delta U$), and the scales over which streams interact ($\Delta r$):
% \begin{align}
%     \Csh = \frac{\Delta U}{U} \frac{r}{\Delta r} .
% \end{align}
% The shear driving is assumed to be at the correlation length scale and thus makes no contribution in the $\lambda$ equation \citep{BreechEA05}.

In summary, 
the TTM parameters that we constrain are $\alpha$, $\beta$, $\sigma_D$, $f_D$, $\Csh$, and the BCs at 0.17\si{AU}
($Z^2_{\init}$, $\lambda_{\init}$, $\sigma_{c,\init}$, and $T_{\init}$).\footnote{In general, 
   Bayesian analysis can be used to constrain any parameter in a TTM.}   
As a starting case 
  %% for our analysis 
we use 
\response{what we will refer to hereafter as our ``nominal'' estimates for}
 $ \sigma_D = -1/3 $, 
 $  f_D = 0.25$ and 
 $ \Csh = 1.5$
(as used in \cite{BreechEA08} for example), 
along with the BCs from the inner-most observations,
and apply Bayesian analysis to just the $\alpha$ and $\beta$ parameters
  (\autoref{sec:ab_analysis}).
Following that, we perform extended analyses that also allow 
  %% variation 
determination of posterior distributions for
    $\sigma_D$, $f_D$, $\Csh$, and the BCs  
  (\autoref{sec:extended_analysis}).

%\red{... say some more about searching over these 3 params, plus BCs, plus $\alpha,\beta$}

%\red{Really, we should also be constraining these source parameters $f_D$ and $\Csh$. This should require minimal effort.} \textcolor{olive}{$f_{D}$ and $\Csh$ can be added to the anaylsis without any modification, essentially meaning the log-evidence for this model is able to be compared to previous results}. \textcolor{teal}{I have added $f_{D}$ and $\Csh$ to the analysis. We obtain $f_{D} \approx 0.25$, and $\Csh \approx 1.27$. However, this comes at the more significant cost of $\alpha_{2D}$ and $\beta_{2D}$. This was with the $\sigma_{D}$ analysis, so the log-evidence is not able to be compared to the previous results.} \textcolor{blue}{Tulasi made a wise comment, that I should leave this analysis for another time.}

\begin{table}
\centering
\begin{tabular}{c c}
\hline
\hline
Quantity & Value\\
\hline
\hline
$ U$                  &  \qty{400}{km/s} \\
$ \sigma_D$           &    $-1/3$        \\
\hline
%% $ V_{A,0}$            & \qty{400}{km/s}  \\
%% $ \Omega_\text{sun}$  & \qty{2.9e-6}{rad/s}\\
%% $ r_{A}$             & $10 \, R_{\odot}$\\
   %% $ \theta$        & $\pi/2$ \\
%% \hline
$ \Csh$              & 1.5\\
$ f_D$               & $0.25$       \\
$ L_{\text{cav}}$    & \qty{8}{AU} \\
$ n^{\infty}_{H}$               & \qty{0.1}{cm^{-3}} \\
$ n^{1\si{AU}}_{SW}$            & \qty{5}{cm^{-3}} \\
$ \tau_{\text{ion}}^{1\si{AU}}$ & \qty{1e6}{s} \\
\hline
\end{tabular}
 \caption{Some TTM parameters common to the 3D and 2D isotropic models. Parameters 
 %% in the middle section are associated with the Parker spiral, and those 
 in the lower section are associated with forcing
   and are as used in \cite{BreechEA08}. \response{We refer to these values as the ``nominal'' estimates.} 
 %% $\Omega_\text{sun}$ is the solar rotation rate.
}
\label{tab:common_model_parameters}
\end{table}

    \subsection{Observational Data}
    \label{sec-obs_data}
The data we assess against is obtained
  % \footnote{The 
  %  datasets we use herein were kindly provided by 
  %   \cite{AdhikariEA15, AdhikariEA21-xport} and \cite{SmithEA06-pi}.} 
    %% We are in the process of creating a newer version of the dataset utilizing significantly more data that have been acquired by \emph{PSP} and \emph{SolO}.}
from \emph{in situ} observations of the solar wind.
   %%For our purposes 
The relevant quantities from the datasets are the 
(traced) rms \elsasser{} energies $Z^2_\pm$, 
their correlation lengths $\lambda_{\pm}$, 
the normalized cross helicity $\sigma_c$, 
the normalized energy difference $\sigma_D$,
and the proton temperature
$T_p \equiv T$ (which is the temperature present in \autoref{eq:model-temp}).
In the inner heliosphere we make use of \emph{Parker Solar Probe} (PSP) measurements made between 0.17--0.6\,\si{AU}
      \citep{AdhikariEA21-xport}.
In  
the outer heliosphere, temperature measurements \citep{SmithEA06-pi} 
and energy and correlation length measurements \citep{AdhikariEA15}
are from the \emph{Voyager}~2 mission. 
To connect the observational values with the TTM variables, we rely on
the usual definitions of total energy
 $ Z^2 = \bracket{Z^2_+ + Z^2_-}/2 $,
correlation length
 $ \lambda = \bracket{\lambda_+ + \lambda_-}/2$, 
and normalized cross helicity
 $ \sigma_c = (Z_+^2 - Z_-^2) / (Z_+^2 + Z_-^2) $.

  %%  Using observational data is a typical method for obtaining the inner boundary values for transport models. When the boundary values are unknown, then a range of values can be used to show a boundary    \red{??? ...}of reasonable solutions available \citep{BreechEA08}.

  %% \red{$T$ at larger $r$ is from Smith}

%We perform a combined analysis of these datasets using Bayesian inference. We can also analyze the contribution of these datasets separately, and whether they provide consistent parameter constraints (\autoref{sec-combined}).

   \section{Bayesian Analysis Approach}
    \label{sec:bayes_analysis}

In this section we present an overview of the Bayesian analysis technique and how it is applied to the Solar Wind TTM, equations~\eqref{eq:model-Z2}--\eqref{eq:model-temp}.

For a model $\mathcal{M}$, and a data vector $\mathcal{D}$, we can obtain the probability distributions of model parameters (or sampling parameters) $\Theta$ according to Bayes theorem:
\begin{align}
    \label{eqn:bayes_theorem}
    Pr(\Theta | \mathcal{D},\mathcal{M}) = \frac{Pr(\mathcal{D} | \Theta, \mathcal{M}) Pr(\Theta | \mathcal{M})}{Pr(\mathcal{D} | \mathcal{M})},
\end{align}
where $Pr(\Theta | \mathcal{D}, \mathcal{M}) \equiv P(\Theta)$ is the 
  \emph{posterior distribution}
of the model parameter set. This quantifies the distributions of, and therefore constrains, the sampled parameters $\Theta$ given the data and model.

$Pr(\mathcal{D}|\Theta, \mathcal{M}) \equiv \mathcal{L}(\Theta)$ 
is the 
    \emph{likelihood function} 
for the data. We use the likelihood function given by
\begin{align}
    \label{eqn:likelihood_function}
    \mathcal{L}(\Theta) = (2\pi)^{-N/2} e^{-\chi^2/2} \prod_{i=1}^{N} \sigma_i^{-1} ,
\end{align}
where $N$ is the number of datapoints, $\sigma_i$ is the error associated with each individual datapoint, and
\begin{align}
    \label{eqn:chi2}
    \chi^2 = \sum_{i=1}^{N} \bracket{\frac{y_i - \tilde{y}_i}{\sigma_i}}^2 ,
\end{align}
for observational data\footnote{Since the 
  data spans many orders of magnitude we found 
  it useful to take the base-10 log of the data $y_i$ and the model estimate $\tilde{y}_i$ before calculating the $\chi^2$.  This helps avoid prioritising one set of parameter observations over another. Consequently, although we do not calculate the true $\chi^2$, the quantity we do calculate plays an analogous role.}
$ y_i $
and model estimates
$ \tilde{y}_i $.
The likelihood function requires evaluation based on the model predictions given the sampled parameters. In other words, it describes the probability of observing the data $\mathcal{D}$, using the parameters $\Theta$ with the model $\mathcal{M}$.

$Pr(\Theta | \mathcal{M}) \equiv \pi(\Theta)$ 
is the 
    \emph{prior probability distribution}
for the model parameter set
and can be used to enforce physical knowledge or assumptions into the statistical inferencing. 
In situations where little is known about the prior one might elect to use a uniform distribution for
  $ \pi(\Theta) $.
The particular prior distributions we employ are stated in
  \autoref{sec:results}.

$Pr(\mathcal{D}|\mathcal{M}) \equiv \mathcal{Z}(\mathcal{D})$ 
is the Bayesian 
    \emph{model evidence}:
the probability of observing the data $\mathcal{D}$, given a model, $\mathcal{M}$. This is a normalizing factor, based on \emph{all} possible $\Theta$ values (the space $\Omega_\Theta$) and is given by the multidimensional integral
\begin{align}
\label{eqn-evidence}
    \mathcal{Z}(\mathcal{D}) = \int_{\Omega_\Theta} \mathcal{L}(\Theta) \pi(\Theta)\,\d^D \Theta ,
\end{align}
where $D$ is the dimensionality of the parameter space (\ie{}the number of model parameters).

To perform our analysis we make use of the freely available package \texttt{pymultinest}\footnote{\url{https://github.com/JohannesBuchner/PyMultiNest}} \citep{Feroz.etal09, Buchner16}, a multimodal nested sampling algorithm (discussed further in \autoref{appendix:mcmc}). This is a Bayesian inference tool that calculates the evidence and produces posterior samples from distributions that may be multimodal, or contain complex degeneracies in high dimensional data.

Currently, we do not have measurements (or estimates) for the error of the datapoints $\sigma_i$. We therefore assume the error is constant and the same for all datapoints (\ie{}$\sigma_i = k$ is constant), and obtain an unnormalized likelihood function by setting $k = 1$. 
Instrumental errors are assumed to have minimal effect on the variables $Z^2$, $\lambda$, $\sigma_c$, $\sigma_D$, and $T$ due to the considerable averaging of data that is performed in obtaining binned radial estimates. 
Ideally, we would have error bars for each variable in each radial bin, based on variation of repeated samplings, or inherent in the randomness of turbulence in the solar wind. 
Overall, more work needs to be done to obtain statistical errors, based on variations over repeated/averaged samplings of solar wind data for all relevant variables represented in the models. 
For example, recent work \citep{CuestaEA22-ec}
reporting the standard deviations for 
binned correlation lengths, $\lambda$, could be incorporated in future Bayesian analyses.

For each selection of parameters $\Theta$ (such as $\alpha$, $\beta$, \etc{}), the Bayesian analysis calls the numerical setup (\autoref{sec-numerical_setup}), which solves the model system and returns the radial solutions for variables $Z^2(r)$, $\lambda(r)$, $\sigma_c(r)$, $T(r)$, and if configured, constant $\sigma_D$\footnote{To keep 
  the number of degrees of freedom the same across our different analyses  
  we need to ensure that the number of elements in the $\chi^2$ summation is always the same. We therefore include the $\sigma_D(r) = \mathrm{const}$ comparison to $\sigma_D$ observations in all analyses, including ones where we 
  set $\sigma_D = -1/3$.} 
  (see eq.~\eqref{eq:dsigDdr}). 
These numerical solutions ($\tilde{y}_i$) are then used in tandem with the observational data ($y_i$)
to calculate the $\chi^2$ values. 
The \texttt{pymultinest} code runs until the evidence estimate has converged, and yields the samples of the posterior distribution, with their weights, based on the likelihood and prior values. Using these samples, we can generate 
estimates for the posterior distributions
and also calculate confidence intervals about the means of the distributions of solutions contained in the posterior samples. 
The obtained posterior distributions are ($N$-dimensional) joint posteriors, and pairs of dimensions can be visualized using two-dimensional contours. 
Some specific 2D joint posterior distributions are discussed in \autoref{sec:discussion}. 
In  cases where the sampled parameters are weakly correlated, it can be appropriate to view just the 1D distributions.

%The triangle plots show the posterior distributions of the sampled parameters, displaying the range of values, the parameters are allowed for the model, given the data. The contour plots show the standard deviations from the solutions in the $3\,\sigma$ interval of the posterior distribution.

In order to decide on an appropriate model to use, it is helpful to compare different models with statistical rigour.
  %% confidently determine the best performing model. 
This can be achieved by evaluating the model evidence 
    $\mathcal{Z}(\mathcal{D}) $; 
  see \autoref{eqn-evidence}. 
Since $\mathcal{Z}(\mathcal{D})$ is the integral of the prior-weighted likelihood over the entire parameter space, it describes the overall effectiveness of the model. 
The 
 %% model evidence 
  $ \mathcal{Z}(\mathcal{D}) $ 
protects against overfitting by penalizing posteriors that cover large fractions of the prior space, and naturally favours simpler models, unless the likelihood is vastly improved. 
Two models, 1 and 2,
may be compared using the Bayes factor, which is essentially
the ratio of their model evidences;
the model with the largest 
  $ \mathcal{Z}(\mathcal{D}) $ 
is most favoured. 
For large values of the model evidences it may be more practical to instead work with
  $ \ln \mathcal{Z}_1 - \ln \mathcal{Z}_2 $. 
Note that since we have essentially assumed no error in the observations, the model evidences are not a completely fair description; however, since the Bayes factor is the ratio of two evidences with the same error assumptions, the results should be defensible.

    \subsection{Numerical Setup}
    \label{sec-numerical_setup}
    
We solve the system 
 \eqref{eq:model-Z2}--\eqref{eq:model-temp}
numerically using LSODA \citep{LSODE} 
for $r$ from 0.17--80\,\si{AU}\footnote{\response{The 
  termination shock has been observed at $r \gtrsim 84$\,\si{AU} 
  \citep[e.g.,][]{McComasEA19} 
  where, there and beyond, our TTM is unlikely to be applicable since it does not include interaction of the solar wind with  the local interstellar medium.
  Note that the choice of $r_\text{max}$ = 80\,\si{AU} for the maximum radius is essentially a cosmetic one. Provided $r_\text{max}$ is greater than the distance of the last observational data point it has no effect on the Bayesian results. 
  This is because the $\chi^2$ values (\autoref{eqn:chi2}) are calculated by comparing the data and the model only at the distances where we have observational values (we use data up to 75\,\si{AU}).}}, 
with a minimum step size of $ 1 \, \si{km} $.
The Python \texttt{scipy} implementation of LSODA, 
    \texttt{solve\_ivp()}, %\footnote{\url{https://github.com/jacobwilliams/odepack}} 
supports an automatic halting of solving the system.  This is an advantageous feature as we may encounter combinations of sampled parameters for which the TTM is not easily solvable. In such cases, we flag a bad parameter selection for \texttt{multinest}.

%\footnote{\url{https://scipy.org/}}

%We use initial conditions $Z^2_{\init}$, $\lambda_{\init}$, $\sigma_{c,\init}$, and $T_{\init}$ for a given choice of constant parameters $\alpha$, $\beta$, and the constants stated in \autoref{tab:common_model_parameters}.

%%    \section{Bayesian Analysis Results}
    \section{Results}
    \label{sec:results}

\begin{figure*}
\centering
\makebox[0cm]{\includegraphics[scale=1]{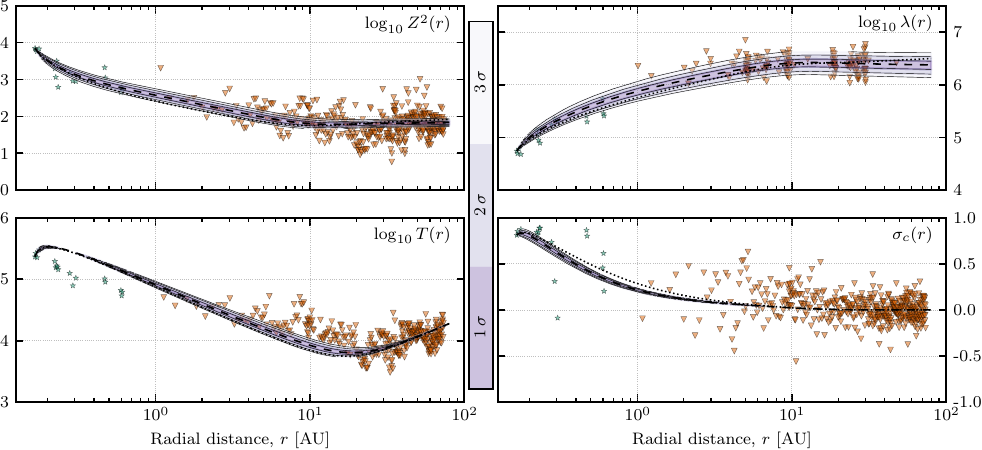}}
   \caption{Solutions of the $\alpha$-$\beta$ Bayesian analysis. The posterior sampled means for the 2D and 3D TTMs are plotted as dashed and dotted black lines respectively. 
   Solid lines depict the $1$, $2$, and $3$ $\sigma$ confidence interval contours in the means of the 2D TTM solutions.  The latter are determined using the sampled posterior distributions for the (2D TTM) $\alpha$-$\beta$ analysis. 
   Green stars indicate \emph{PSP} data points, and orange triangles \emph{Voyager} 2 data points. Units are as in \autoref{tab:observation_initial_values}.}
  %.. NOTE: technical distinction between posterior predictive distribution (which requires error bars) and confidence intervals 
 \label{fig:ab_contour}
\end{figure*}

We carry out several Bayesian/\texttt{multinest} analyses of the available models, beginning with the smaller analyses that constrain only the 
vKH $\alpha$--$\beta$ parameters, 
named the $\alpha$--$\beta$ analysis. 
  %% This showcases the simplest scenario of constraining the unknown, constant $\alpha$-$\beta$ parameters. 
Next, we perform an analysis that includes most of the other model parameters which are constant (but not precisely determined), specifically $\sigma_D$, $f_D$, and $\Csh$.
This is referred to as the 
$\alpha$-$\beta$-$\sigma_D$-$f_D$-$\Csh$ analysis. 
Analyses with respect to the BCs  for the four transported quantities 
  ($Z^2_{\init}$, $\lambda_{\init}$, 
   $\sigma_{c,\init}$, and $T_{\init}$), 
treating them as parameters with distributions to constrain,
are also performed.
These can be useful when the boundary values are unknown and/or display observational scatter.
This latter approach may also capture more of the scatter present in the larger radii solar wind observations. 

Two other analyses are also executed.  The first showcases constraining the model parameters using split datasets.
This enables examination of how the different datasets contribute to the constraints on the sampled parameters.  
Lastly, we investigate the effects of removing the PI driving term from the $\lambda$ evolution equation.

    \subsection[alpha--beta analysis]{$\alpha$-$\beta$ Analysis}
    \label{sec:ab_analysis}

\begin{table}
\centering
\begin{tabular}{c c}
\hline
\hline
%% Inner Boundary Values & Values\\
Quantity & Value \\
\hline
\hline
%$Z^2_{\init}$ & $7134$\,\si{(km/s)^2}\\
%$\lambda_{\init}$ & $165000$\,\si{km}\\
%$\sigma_{c,\init}$ & $0.894$\\
%$T_{\init}$ & \si{K}\\
%$Z^2_{\init}$ & $7,039.49$\,\si{(km/s)^2}\\
$Z^2_{\init}$ & $7,040$\,\si{(km/s)^2}\\
%$\lambda_{\init}$ & $55,351.21$\,\si{km}\\
$\lambda_{\init}$ & $55,400$\,\si{km}\\
$\sigma_{c,\init}$ & $0.81$\\
%$T_{\init}$ & $232,313.06$\,\si{K}\\
$T_{\init}$ & $232,000$\,\si{K}\\
\hline
\end{tabular}
  \caption{Inner boundary values 
    employed when these parameters are assigned $\delta$-function priors (\ie{}kept fixed) in the Bayesian analysis. 
    Values are approximately those from PSP data presented in
       \cite{AdhikariEA15}.
    %\red{I don't think we need to use this many sig figs. 
    % 3 or maybe 4 sig fig would be good (and more readable)}
    }
\label{tab:observation_initial_values}
\end{table}

In this section, we perform what are our smallest analyses for the 2D and 3D TTMs, searching only over $\alpha$ and $\beta$. 
As priors for $\alpha$ and $\beta$, we use the uniform distributions stated in \autoref{tab:alpha_beta_prior},
where the intention is to cover as much of the parameter space as possible without influencing the posterior distribution, that is, we are keeping the priors as uninformative as possible. 
The inner boundary values, $Z^2_{\init}$, $\lambda_{\init}$, $\sigma_{c,\init}$, $T_{\init}$,
are fixed and based on PSP observations at
  $r_0 \approx 0.17\,\si{AU} $,
  as reported in
  \cite{AdhikariEA15}.
See \autoref{tab:observation_initial_values}.%\footnote{Recall that there is an important technical point regarding the use of $\sigma_D = const.$ approximation. Although we are maintaining $\sigma_D \approx -1/3$ in this subsection, it is important to include the calculation of the constant $\sigma_D(r)$ (i.e., \autoref{eq:dsigDdr}) in \autoref{eqn:chi2} in order for the log-evidences in this section to be comparable to those of the next section.}

%As ``standard'' BCs
%for $Z^2$, $\lambda$, $\sigma_c$, and $T$ 
%we use observational data at $0.17\,\si{AU}$; 
%see \autoref{tab:observation_initial_values}.
%\red{Better to move this last sentence, and Table 2, about std BCs to Sec 4.1?} 

%There is a technical point regarding use of the $\sigma_D = const.$ approximation.
% ....\\
%Although we are maintaining $\sigma_D \approx -1/3$, it is important to include the calculation of %the constant $\sigma_D$ 
%  (i.e., \autoref{eq:dsigDdr})
%in the likelihood evaluation in order for the log-evidences in this section to be comparable to %those of the next section.

\begin{table}[!htb]
\centering
\begin{tabular}{c c}
\hline
\hline
Parameter & Prior distribution \\
\hline
\hline
$\alpha$ & $\mathcal{U}[0.01, 2]$\\
$\beta$ & $\mathcal{U}[0.01, 2]$\\
\hline
\end{tabular}
\caption{Prior distributions of the $\alpha$ and $\beta$ vKH parameters used in the Bayesian analysis of the 2D and 3D TTMs. $\mathcal{U}[\cdot, \cdot]$ denotes a uniform distribution with the arguments indicating the domain boundaries.}
\label{tab:alpha_beta_prior}
\end{table}

If we had error estimates for the observational data points, we would approximate a 
    \emph{posterior predictive distribution} (PPD), 
and use it to understand how well the model fits the data. 
A PPD 
  %% predictive posterior distribution 
describes the distribution of possible unobserved values. That is, it takes into account the uncertainties of the mean of the posterior distribution for the TTM parameters, as well as the uncertainties in the observations to calculate the probabilities of future observations. A PPD would be a rigorous way of using a TTM to predict future measurements of the solar wind 
 %% at a given 
  as a function of
heliocentric radius.
%\blue{Since the aim of the PPD is to predict future observations it naturally accounts for the error of the }
%\red{With the errors, the confidence intervals would take account of the uncertainties in the data, and therefore characterise the PPD. (ask Yvette)...}

Since, however, we do not have estimates for the errors in the data points (we are neglecting all forms of uncertainty in the observations)
the PPDs are not able to be calculated.
In
\autoref{fig:ab_contour} 
we instead display the predicted radial distribution of $Z^2(r)$, $\lambda(r)$, $T(r)$, and $\sigma_c(r)$ based on the mean of the estimated posterior distribution, 
 $ P(\Theta) $,
of the constrained parameters $\Theta$ in the TTM. 
The uncertainties associated with these predictions\footnote{We emphasize 
again that these confidence interval uncertainties do no necessarily capture the full uncertainty associated with the observations.}
are characterized by the confidence intervals and displayed using contours of the 1, 2, and 3 $\sigma$ deviations. These contours are calculated by solving the TTM for each parameter's posterior sample. We can then calculate the mean, and confidence intervals, given the range of these TTM solutions provided for each radial distance. \emph{If} the uncertainties in the parameters $\Theta$ (as described by the estimated posterior distributions) are expected to describe the scatter in the observations, we would expect the confidence intervals to cover most (if not all) of the observed data points.

From \autoref{fig:ab_contour}
it is evident that overall the solutions match the data reasonably well, although 
the scatter in the observations is not completely captured 
 (except for $\lambda(r)$, which describes the data well). 
The qualitative behaviour of
$T(r)$ accords with the expected evolution: slower than the adiabatic $r^{-4/3}$ profile, with an increase at large radial distances due to the PI driving. 
However, the $T(r)$ solutions show an unphysical increase in temperature near the inner boundary ($ 0.17 < r < 0.2 $).  Furthermore, they exhibit values systematically larger than the \emph{PSP} observational data at small radial distances, and smaller than the bulk of \emph{Voyager}~2 data at intermediate radial distances ($r \sim 10\,\si{AU}$). 
Notably, the
$\sigma_c(r)$ solutions are all very similar, lying almost on top of each other, 
and quickly converge towards $\sigma_c = 0$ with increasing $r$. This is distinctly different from the observational data that is scattered between approximately $\pm 0.5$ beyond $r \approx 1\,\si{AU}$. This scatter is expected to be from variability in the solar wind turbulence rather than due to the effects of observational noise.

  \subsection[Analyses including all parameters]{Analyses including $\sigma_D$, $f_D$, $\Csh$, and Boundary Conditions}
     \label{sec:extended_analysis}

In the previous subsection we carried out Bayesian analysis allowing just the $\alpha$ and $\beta$ TTM parameters to vary.  Here we extend the analysis so that all five of the adjustable parameters in the TTM, plus the inner BCs are allowed to vary. In addition to this ``full'' analysis we also assess several subcases where only some of the model parameters and/or BCs are analysed over.
The three combinations we investigate  are 
\vspace*{-1.1ex}
\begin{enumerate} \itemsep=-0.7ex
   \item[(i)] $\alpha$-$\beta$-BC,
   \item[(ii)] $\alpha$-$\beta$-$\sigma_D$-$f_D$-$\Csh$,  
   \item[(iii)] $\alpha$-$\beta$-BC-$\sigma_D$-$f_D$-$\Csh$,
 \end{enumerate}
and for each of these the 2D and 3D TTMs are assessed.
The extended analyses require sampling more of the TTM parameters, and therefore more evaluations of $\chi^2$ which increases the code run times. In our case, the analyses are still acceptably quick to compute even when run on a reasonably standard desktop computer (\autoref{appendix:mcmc}).

The $\alpha$-$\beta$-BC case includes the BCs---$Z^2_{\init}$, $\lambda_{\init}$, $\sigma_{c,\init}$, and $T_{\init}$---as parameters to constrain. 
Their 
 % boundary condition 
prior distributions are described in the top section of \autoref{tab:extended_priors}, with the $\alpha$ and $\beta$ priors still given by \autoref{tab:alpha_beta_prior}. 
We choose the BC priors to be as uninformative as possible, namely uniform for $Z^2_{\init}$, $\lambda_{\init}$, and $\sigma_{c,\init}$, and a log-normal distribution for $T_{\init}$. 
  The mean and variance for the $T_{\init}$ prior were chosen to accord with the inner most \emph{Helios} observations discussed in \cite{HellingerEA13}. 
We remark that when a uniform prior for $T_{\init}$ was investigated, we found the resulting posterior distribution had a large tail extending towards unreasonably small values.
Allowing the BCs to have distributions makes some allowance for the non-steady nature of the solar wind observations. 
For example, observed values at $5\,\si{AU}$ and $10\,\si{AU}$ typically will not correspond to the same parcel of solar wind transported from 
    $ r = \init\,\si{AU} $.

%Constraining the initial conditions potentially acknowledges the fact that later radii observations will not necessarily correspond to the same initial conditions in the outer heliosphere (allowing variability in the steady-state assumption). \red{I am struggling to word what I want here.}. \YP{How about: Allowing the initial conditions to vary acknowledges the fact that observations at larger radii could be sampling solar wind resulting from different initial conditions to the observations at smaller radii; i.e.\ there could be deviations from the steady-state assumption.}

For the
$\alpha$-$\beta$-$\sigma_D$-$f_D$-$\Csh$ case the analysis is 
essentially the same as that used for the $\alpha$-$\beta$ case discussed in \autoref{sec:ab_analysis}, 
but with the inclusion of the other adjustable parameters: 
 %% (assumed) constant 
  $\sigma_D$, $f_D$, and $\Csh$. 
We employ uniform priors for these, as listed in the second section of \autoref{tab:extended_priors} (and \autoref{tab:alpha_beta_prior}), 
once again keeping them as nonrestrictive as possible within the range of reasonable expected values. 
Note that the uniform prior for $\Csh$ is extended compared to the 0--2 range of values described by \cite{BreechEA08}. 

The third case, 
$\alpha$-$\beta$-BC-$\sigma_D$-$f_D$-$\Csh$, 
contains all the parameters described thus far, using the priors already stated (\autoref{tab:alpha_beta_prior} and \autoref{tab:extended_priors}). 
%% and is the most computationally expensive (see \autoref{mcmc:tab:time} for the run times on a desktop computer). 

\begin{table}
\centering
\begin{tabular}{c l}
\hline
\hline
Quantity & Prior distribution \\
\hline
\hline
$Z^2_{\init}$  & $\mathcal{U}[10^2\,\si{(km/s)^2}, 10^5\,\si{(km/s)^2}]$ \\
$\lambda_{\init}$    & $\mathcal{U}[10^3\,\si{km}, 10^7\,\si{km}]$\\
$\sigma_{c,\init}$ & $\mathcal{U}[0, 1]$\\
$T_{\init}$ & $\ln\mathcal{N}(2.5 \times 10^5\,\si{K}, 0.5)$\\
\hline
$\sigma_D$ & $\mathcal{U}[-1, 1]$\\
$f_{D}$ & $\mathcal{U}[0.01, 1]$\\
$\Csh$ & $\mathcal{U}[0.01, 8]$\\
\hline
\end{tabular}
  \caption{Prior distributions employed in the analyses that include the 
  BCs (top half) 
  %% $Z^2_{\init}$, $\lambda_{\init}$, $\sigma_{c,\init}$, and $T_{\init}$, 
  and the constant TTM parameters (lower half). 
  %% $\sigma_D$, $f_D$, and $\Csh$. 
  The priors for $\alpha$ and $\beta$ are stated in \autoref{tab:alpha_beta_prior}. $\mathcal{U}[\cdot, \cdot]$ denotes a uniform distribution with the arguments indicating the domain boundaries, and $\ln \mathcal{N}(\mu, \sigma)$ denotes a log-normal distribution with location parameter $\mu$ and logarithm of scale parameter $\sigma$.}
  %\green{The log-normal distribution $\ln\mathcal{N}(\mu, \sigma)$ is described by the following probability density function $f(x; \mu, \sigma) = \frac{1}{x \sigma \sqrt{2 \pi}} \e^{- \frac{\ln(x/\mu)^2}{2 \sigma^2}}$ where $\mu$ is the location parameter and $\sigma$ is the logarithm of scale parameter.}}
\label{tab:extended_priors}
\end{table}

\begin{figure*}
\centering
\makebox[0cm]{\includegraphics[scale=1]{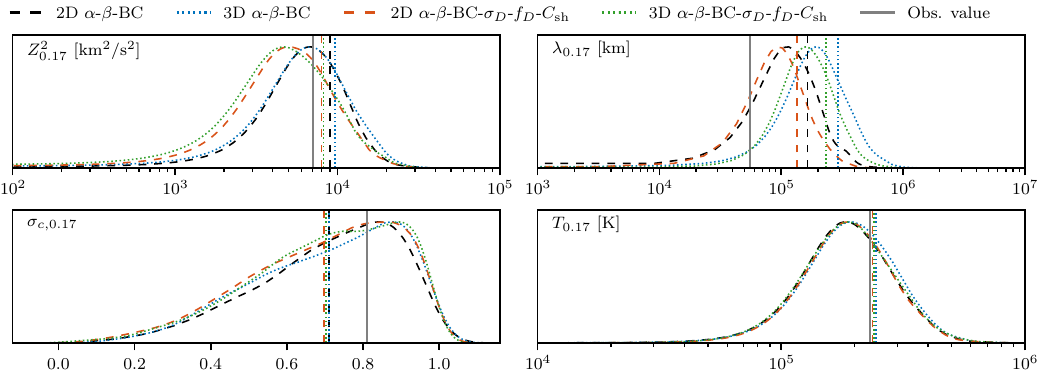}}
  \caption{1D posterior distributions of the 2D (dashed) and 3D (dotted) models of the $\alpha$-$\beta$-BC and $\alpha$-$\beta$-BC-$\sigma_D$-$f_D$-$\Csh$ analyses. Vertical lines with the same linestyle and colour represent the mean value of the sampled posterior distributions. For comparison the vertical gray line is the \emph{PSP} inner most observation, used in the analyses that do not constrain these variables ($\alpha$-$\beta$, and $\alpha$-$\beta$-$\sigma_D$-$f_D$-$\Csh$); see \autoref{tab:observation_initial_values}.
  %\red{Better for the grey line legend to be `standard value' or 'Observational value'?}
  }
\label{fig:initial_distributions}
\end{figure*}
\autoref{fig:initial_distributions} 
displays the 1D posterior distributions for the BCs obtained from both 
the $\alpha$-$\beta$-BC analysis
and the $\alpha$-$\beta$-BC-$\sigma_D$-$f_D$-$\Csh$ analysis, 
cases (i) and (iii).
Results are shown for the 2D and 3D TTMs. 
The vertical lines show the mean value of the respective posterior distributions.  These mean values may be treated as corresponding to the ``best'' solutions. 
Considering the visually simplest distribution first, the $T_{\init}$ (bottom right) posterior distributions are essentially identical, with mean values that align well with the \emph{PSP} observation. 
Similarly, the $\sigma_{c,\init}$ (bottom left) posterior distributions for the different analyses are similar, resulting in mean values of $\sigma_{c,\init} \approx 0.7$, 
slightly smaller than the observational value of $\sigma_{c,\init} \approx 0.8$. 
For a given case---(i) or (iii)---the 
    $Z^2_{\init}$ (top left) 
1D posterior distributions are very similar for the two TTMs (i.e., the 2D and 3D models). 
The mean is slightly larger for the $\alpha$-$\beta$-BC analyses. 
Although the $Z^2_{\init}$ \emph{PSP} observation is smaller than the mean sampled posterior values, the posterior distributions are consistent with the observational value. 
The situation for
$\lambda_{\init}$ (top right) is switched relative to that for $Z^2_{\init}$,
with the posteriors for the 2D TTM being similar across cases (rather than across TTM models), and likewise for the 3D TTM posteriors, although with these shifted to larger values.
 %%  difference in the posterior distributions of the 2D and 3D models for the $\alpha$-$\beta$-BC and $\alpha$-$\beta$-BC-$\sigma_D$-$f_D$-$\Csh$ analyses, with the 3D model favouring the larger values. 
The resulting mean of the sampled posterior distribution for the 3D $\alpha$-$\beta$-BC analysis is almost an order of magnitude larger than the \emph{PSP} observation. The observed value is still consistent with the $\lambda_{\init}$ posterior distributions, but only marginally so for the 3D models.

\begin{figure*}
\centering
\makebox[0cm]{\includegraphics[scale=1]{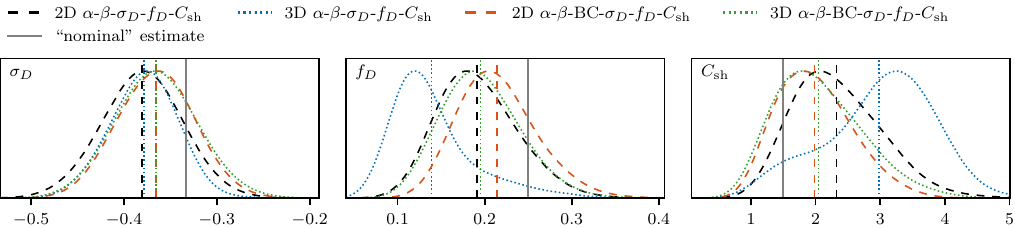}}
\caption{The 1D posterior distributions of $\sigma_D$ (left), $f_D$ (middle), and $\Csh$ (right) for the 2D and 3D TTMs (dashed and dotted) of the $\alpha$-$\beta$-$\sigma_D$-$f_D$-$\Csh$ and $\alpha$-$\beta$-BC-$\sigma_D$-$f_D$-$\Csh$ analyses. The vertical lines represent the mean value of the respective plotted sampled posterior distributions. For comparison, the vertical gray lines are the assumed values for the analyses that do not constrain these variables ($\alpha$-$\beta$, and $\alpha$-$\beta$-BC); see \autoref{tab:common_model_parameters}.}
\label{fig:sigd_fd_csh_distributions}
\end{figure*}
  \autoref{fig:sigd_fd_csh_distributions} presents 1D posterior distributions 
from cases (ii) $\alpha$-$\beta$-$\sigma_D$-$f_D$-$\Csh$
and (iii) $\alpha$-$\beta$-BC-$\sigma_D$-$f_D$-$\Csh $,
once again for both the 2D and 3D TTMs so that there are four different distributions on each panel.
Shown are the posteriors 
for $\sigma_D$, $f_D$, and $\Csh$.

For $\sigma_D$ the four posterior distributions are quite similar and consistent with the commonly used (and observationally based) value of  $\sigma_D \approx -1/3$ \citep{PerriBalogh10-sigc}. The $\alpha$-$\beta$-$\sigma_D$-$f_D$-$\Csh$ analysis has mean values of $\sigma_D \approx -0.38$, and the $\alpha$-$\beta$-BC-$\sigma_D$-$f_D$-$\Csh$ has mean values of $\sigma_D \approx -0.36$. 
Moving on to the $f_D$ distributions, 
these are similar for the (2D and 3D) $\alpha$-$\beta$-BC-$\sigma_D$-$f_D$-$\Csh$ analysis and just the 2D $\alpha$-$\beta$-$\sigma_D$-$f_D$-$\Csh$ analysis. 
Most of the mean values are 
$f_D \approx 0.2$,
slightly smaller than the \response{nominal} value of $f_D \approx 0.25$. 
The exception is for the 3D $\alpha$-$\beta$-$\sigma_D$-$f_D$-$\Csh$ analysis, for which the posterior distribution and  
the mean $f_D \approx 0.12$
are both quantitatively distinct from the other cases.

Examining the
 $\Csh$ posterior distributions we see that they extend well past the 
$\Csh \in \mathcal{U}[0, 2]$
range assumed in \cite{BreechEA08},
motivating our use of an extended range for the prior. 
Without extending the prior, the obtained posterior distributions would be unnaturally constrained against the $\Csh=2$ wall. 
All of the mean values are larger than the \response{nominal} value of 
 $ \Csh \approx 1.5 $ 
   \response{(\autoref{tab:common_model_parameters})}. 
For the 2D and 3D $\alpha$-$\beta$-BC-$\sigma_D$-$f_D$-$\Csh$ analyses the means are $\Csh \approx 2$, values which are not in tension with $\Csh = 1.5$ 
given the widths of the distributions. 
Once again, the 3D $\alpha$-$\beta$-$\sigma_D$-$f_D$-$\Csh$ analysis shows a distinct distribution compared to the other results, and an even larger mean value of $\Csh \approx 3$. The extended range of the $\Csh$ prior and posterior distributions are considered further in \autoref{sec:split_analysis}.

\begin{figure*}
\centering
\makebox[0cm]{\includegraphics[scale=1]{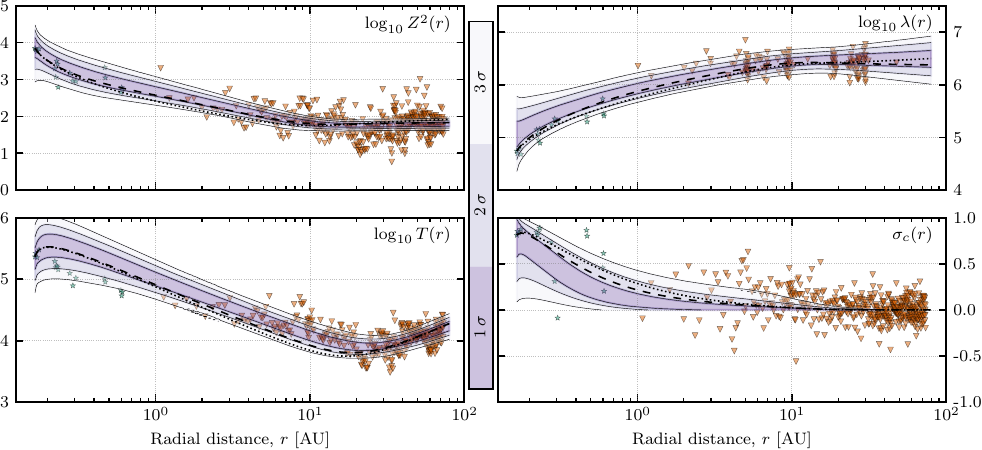}}
\caption{Solutions for the 2D $\alpha$-$\beta$-BC-$\sigma_D$-$f_D$-$\Csh$ analysis. The contours represent the confidence intervals of the means in the solutions obtained by the sampled posterior distributions. Green stars show the \emph{PSP} data, and the orange triangles the \emph{Voyager}~2 data. The 2D and 3D TTM posterior sampled means of the $\alpha$-$\beta$ analysis are plotted with dashed, and dotted black lines respectively. Units are as in \autoref{tab:observation_initial_values}.}
\label{fig:all_contour}
\end{figure*}

%\YP{I don't like the word "variability" as you've used it here.  To me, variability means something actually physically changing in the system.  You're talking about the uncertainty in the model solution increasing, which is just due to the greater freedom you've allowed the parameters to have.  The fact that the model distribution doesn't cover all the datapoints is not necessarily a bad thing - think about the sqrt(N) thing here as well.  Your $\sigma_c$ values at large radii just look evenly scattered around 0, so the difference from the model is probably just due to individual errorbars rather than the model not capturing the variation.}

   \autoref{fig:all_contour} 
       %% (akin to \autoref{fig:ab_contour})
shows  confidence intervals for the 2D case (iii)
 $ \alpha$-$\beta$-BC-$\sigma_D$-$f_D$-$\Csh$ analysis. 
For comparison, the solutions shown in \autoref{fig:ab_contour} 
(\ie{} using the mean sampled posterior values for the 2D and 3D $\alpha$-$\beta$ analysis) 
are also included. 
See 
 %% \autoref{fig:sigd_contour_all} in
Appendix~\ref{appendix:mcmc} for the confidence intervals of $\sigma_D$ for this same analysis case.

We can see that the inclusion of the additional parameters (and the BCs) in the Bayesian analysis 
leads to confidence intervals that span a greater range of values.
%% creates much more uncertainty due to the greater freedom, as indicated by the wider confidence intervals.
In other words %Consequently
 there is more uncertainty associated with the ``best'' TTM solution.
%% This indicates the Bayesian analysis is \red{more uncertain in the TTM solutions ??}.
%\red{Although this means that more of the data is covered by a confidence interval it comes at the cost of greater uncertainty.}
If the confidence interval contours were to cover the majority (or all) of the scatter in the observations, this greater uncertainty could be considered advantageous since it supports explaining the scatter in the data as variability in the solar wind caused by non-steady state behaviour.

The range in the $Z^2(r)$, $\lambda(r)$, $T(r)$, and $\sigma_c(r)$ solutions has increased, compared to the $\alpha$-$\beta$ analysis,
allowing them to capture much more of the observational scatter. 
We still see the unphysical increase in $T(r)$ close to the inner boundary, and indeed it is present in all of the $T(r)$ confidence intervals.
Additionally, 
the $T(r)$ confidence intervals typically cover a wider range than the PSP observational data does. However, on the plus side, compared to the $\alpha$-$\beta$ analysis, $T(r)$ captures much more of the observed \emph{Voyager} 2 observations. For $\sigma_c$ we see that the confidence intervals are wide at smaller radial distances, but converge to $\sigma_c = 0$ with increasing distance. The scatter in the $\sigma_c$ observations is not well captured.

Note that the mean solutions from the 2D $\alpha$-$\beta$ analysis tend to lie within the 1$\sigma$ or 2$\sigma$ confidence intervals for the $\alpha$-$\beta$-BC-$\sigma_D$-$f_D$-$\Csh$ analysis.

%at large and intermediate radial distances respectively. Whereas the initial conditions

%Initial conditions allows for variation in the inner heliosphere

As the confidence intervals indicate, the non-steady state and variation of parameters allowed by the 2D $\alpha$-$\beta$-BC-$\sigma_D$-$f_D$-$\Csh$ analysis is able to capture some of the scatter in the observations. However, the additional parameters are not completely sufficient to model all of the scatter in the observations, particularly for $Z^2$ and $\sigma_c$ at large radial distances.

  %% \red{talk to Sean about the variability in $\sigma_c$, what could cause that?}

%The different combinations of parameter analyses 
%(not shown)
%\red{different analyses}

%\autoref{fig:all_contour} shows the posterior predictive distribution plot; described in detail in the previous section (\autoref{fig:ab_contour}), for the 2D $\alpha$-$\beta$-BC-$\sigma_D$-$f_D$-$\Csh$ analysis. For comparison, the solutions shown in \autoref{fig:ab_contour}, \ie{}using the mean sampled posterior values for the 2D and 3D $\alpha$-$\beta$ analysis are included. We can see that the inclusion of the additional parameters (mainly the boundary conditions) in the Bayesian analysis allows for much wider distributions in the predictive posterior plots. The range in the $Z^2$ and $\lambda$ solutions is increased, allowing them to capture much more of the observational scatter than the $\alpha$-$\beta$ analysis. We still see the unphysical increase in the temperature at the start of the evolution. $\sigma_c$ has a wider distribution in the early radial distances, but converges to $\sigma_c = 0$ quickly. It does not capture the scatter in the $\sigma_c$ observations, especially at large radial distances. The mean solutions of the $\alpha$-$\beta$ analyses lie within the $1$, $2$, and $3$ $\sigma$ contours of the predictive posterior distribution.

    \subsection{Split Dataset Analysis}
    \label{sec:split_analysis}

Our observational data comes from two separate spacecraft missions, \emph{PSP} 
($ < 1 \,\si{AU}$) 
and \emph{Voyager}~2 
($ > 1 \,\si{AU}$). 
To check for consistency and biases we repeated the Bayesian analysis using just one dataset at a time and compared the individual results to each other, and to the combined dataset analysis presented in the previous section. 
The results obtained (not shown) are consistent with those discussed in the rest of this section.

Separate dataset analyses could provide independent indications of the effect of the ``inner'' heliosphere data on the whole TTM model and of the ``outer'' heliosphere data on the whole TTM model.
Alternatively, we could investigate splitting the data based on relevant physical differences that apply inside and outside an appropriate distance.  We report on this second option below, choosing the splitting distance to be 
    $ 5\,\si{AU} $.  
This distance is a little smaller than the radius of the PI ionization cavity, $L_\text{cav} = 8\,\si{AU}$, and a little larger than the distance where large-scale stream shear is strongest in the inner heliosphere. %\green{We have also performed this split analysis with respect to the two different datasets ($1 \, \si{AU}$ split) }

We show the split datasets (split at $5\,\si{AU}$) using all of the available parameters \ie{}$\alpha$-$\beta$-BC-$\sigma_D$-$f_D$-$\Csh$ with the 2D model. For both portions of the split ($<5\,\si{AU}$, and $>5\,\si{AU}$), and for the joint dataset (i.e., the 2D $\alpha$-$\beta$-BC-$\sigma_D$-$f_D$-$\Csh$ analysis in \autoref{sec:extended_analysis}), the model is solved from the inner radius $r_0=0.17\,\si{AU}$ to $r_{\text{max}} = 80\,\si{AU}$. The only difference is the data we are including in the likelihood evaluation (\autoref{sec:bayes_analysis}). 
  % We also carried out the analyses with the individual spacecraft datasets
  % (equivalent to splitting at $1\,\si{AU}$). 
  % The results (not shown)
  % are consistent with those obtained from the $ 5 \, \si{AU} $ split.

\begin{figure*}
\centering
\makebox[0cm]{\includegraphics[scale=1]{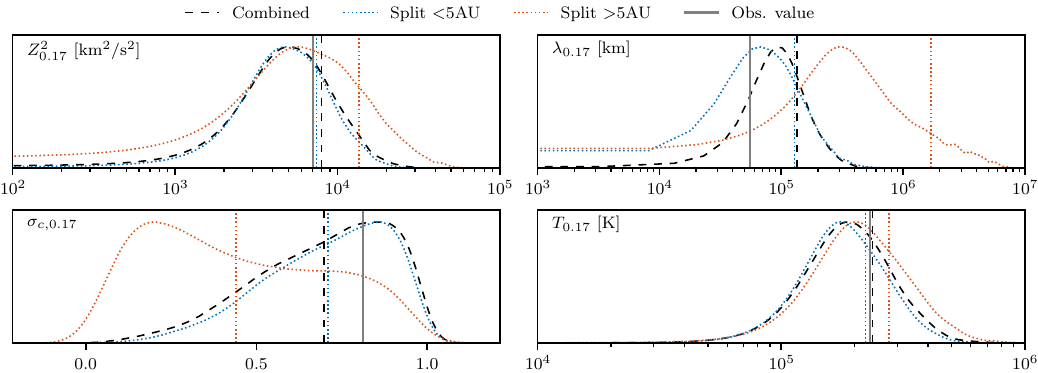}}
  \caption{1D posterior distributions for the 2D model 
  $\alpha$-$\beta$-BC-$\sigma_D$-$f_D$-$\Csh$ 
  analyses using split datasets: 
  $ <5 \,\si{AU}$ (dotted blue), and 
  $ >5 \,\si{AU}$ (dotted orange), and 
  the combined dataset (dashed black). 
  Vertical lines represent the mean value of the sampled posterior distributions. For comparison the vertical gray line is the \emph{PSP} inner most observation, used in the analyses that do not constrain these variables ($\alpha$-$\beta$ and $\alpha$-$\beta$-$\sigma_D$-$f_D$-$\Csh$); see \autoref{tab:observation_initial_values}.}
\label{fig:split_initial_distributions}
\end{figure*}

\autoref{fig:split_initial_distributions} shows the 1D posterior distributions for the BCs  of the $5\,\si{AU}$ split-dataset analyses along with the already presented combined analysis of \autoref{sec:extended_analysis}. We obtain consistent posterior distributions for $T_{\init}$ for both splits. %We see that $T_{\init}$ obtains consistent posterior distributions for both splits of the dataset and the combined analysis, as well as mean values consistent with the observational data. 
For $\sigma_c$, the $<5\,\si{AU}$ analysis is consistent with the combined analysis, whereas the $>5\,\si{AU}$ analysis gives a very unconstrained posterior distribution, favouring the smaller values. This is expected due to the roughly symmetric spread about zero in the \emph{Voyager}~2 $\sigma_c$ dataset for $r>5\,\si{AU}$. It is also expected that the $<5\,\si{AU}$ dataset contributes the most to the joint analysis (evidenced by the identical distributions) because it is the BCs that correspond to the inner heliosphere observations. Similarly, the inner heliosphere analysis constrains $Z^2_{\init}$ (again, identical distributions).
The outer-heliosphere posteriors are also consistent with the inner heliosphere posteriors but with slightly larger mean values.
  %% \red{The outer-heliosphere dataset is also consistent with the inner heliosphere dataset}, 
  %% but with a slightly larger mean value. 
  %% \red{Do we mean consistency of the posteriors (not of the datasets)?... \newline}
For $\lambda_{\init}$, 
the outer-heliosphere posterior is broad, with a mean much larger than those for the inner-heliosphere and the combined analysis. The combined analysis is influenced by both datasets, but predominantly the $<5\,\si{AU}$ dataset. Overall, there is no substantial disagreement between the posteriors for $Z^2_{\init}$, $\lambda_{\init}$, $\sigma_{c,\init}$, and $T_{\init}$ obtained by from the three datasets (combined, $<5\,\si{AU}$, and $>5\,\si{AU}$).

\begin{figure*}
\centering
\makebox[0cm]{\includegraphics[scale=1]{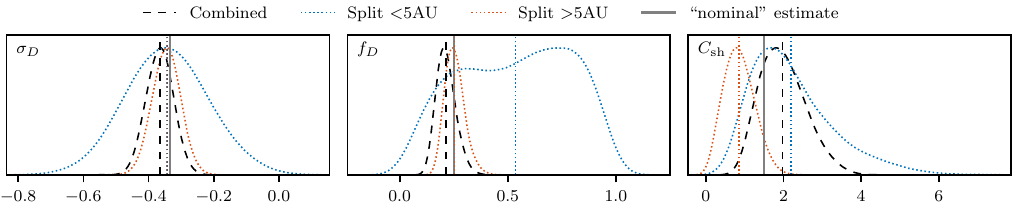}}
  \caption{1D posterior distributions of $\sigma_D$ (left), $f_D$ (middle), and $\Csh$ (right) for the 2D model $\alpha$-$\beta$-BC-$\sigma_D$-$f_D$-$\Csh$ analyses using the: $<5\,\si{AU}$ dataset (dotted blue), the $>5\,\si{AU}$ dataset (dotted orange), and the combined dataset (dashed black). For comparison, the vertical gray lines are the values imposed for the analyses that do not constrain these variables; 
   %% (i.e., $\alpha$-$\beta$, and $\alpha$-$\beta$-BC);
  see \autoref{tab:common_model_parameters}.
  %\red{What does 'Obs value' mean for $f_D$ and $\Csh$?  They're not actually observational values, rather the nominal estimates from \autoref{tab:common_model_parameters}, right? Can we change 'Obs. value' to, say, 'nominal estimate' ?}
  }
\label{fig:split_sigd_fd_csh_distributions}
\end{figure*}

\autoref{fig:split_sigd_fd_csh_distributions} shows the 1D posterior distributions for $\sigma_D$, $f_D$, and $\Csh$. The inner heliosphere dataset for $\sigma_D$ yields a less constrained posterior distribution compared to those for the other two cases which are rather tightly peaked near $ \sigma_D \approx -1/3 $.
This indicates that it is the outer heliosphere dataset that is controlling the behaviour for the combined dataset.
The mean values for the two split datasets are similar and consistent with the combined analysis, and with the assumption of $\sigma_D \approx -1/3$ used
in the simplest ($\alpha$-$\beta$) analysis. 
Expectedly, the inner heliosphere dataset shows an unconstrained result for $f_D$, as the PI driving term is near zero at 
    $r < 5\,\si{AU}$, 
due to the exponential factor in \autoref{eq:Epi-dot}. 
The $\Csh$ TTM parameter constraints are significantly different for the inner and outer heliosphere datasets. The outer-heliosphere posterior is relatively narrow and spans the value range of $[0,2]$, as discussed by \cite{BreechEA08}. The inner-heliosphere posterior is less constrained and extends well past $\Csh = 2$. This is the cause of the combined data analysis results, as introduced in \autoref{sec:extended_analysis}. The inner heliosphere dataset requires a considerably larger value of $\Csh$, and this value predominates in the combined analysis.
This accords with expectations that 
shear driving is stronger in the inner heliosphere, since large-scale velocity gradients tend to decrease with heliocentric distance 
    \citep[e.g.,][]{ZhouMatt90a,BrunoCarbone13}.
  %% Physically, this means that the inner-heliosphere requires much more shear driving than the outer-heliosphere, at least as modelled within these TTMs, 
However, this result also suggests that use of $\Csh=\mathrm{const}$ is not adequate to properly capture the solar wind dynamics over a large range of distances. Although note, these $\Csh$ results are within $\approx 1 \sigma$ of each other so there is no statistically significant disagreement.

Thus, overall, we find no fundamental disagreement between the posterior distributions obtained from the individual datasets and from the combined datasets. The posterior distributions obtained for each split dataset mostly show significant overlap with the posteriors for the combined dataset. This indicates that the Bayesian constraints are not being distorted or biased, given our current TTM, by (known or unknown) observed features of the solar wind. The inclusion of more data correctly serves to improve the Bayesian constraints and does not provide any conflict to our results.
%This indicates that the Bayesian analysis is not being distorted/biased by (known or unknown) features of the solar wind that are modelled in our TTM. 
%Conclude the datasets do not have significant `problems'...
%----------------------------------------------------------

 %% \clearpage

    \subsection[No Pickup Ion Driving of lambda]{No Pickup Ion Driving of $\lambda$}
    \label{sec:no_PI_analysis}

  % This analysis subsection has the same setup as \autoref{sec:ab_analysis} and \ref{sec:extended_analysis}, with priors as given in Tables \ref{tab:alpha_beta_prior} and \ref{tab:extended_priors}.
Equation~\eqref{eq:model-lambda} for the turbulence correlation length 
  $ \lambda $
contains a term involving $ \Epidot$.  
This arises via imposition of the local conservation law 
 $ Z^{2\beta/\alpha}\lambda$ 
 \response{(just for the pickup ion effects)}
and its implication that
\begin{align}
    \Deriv{\lambda}{r}\Bigg|_\text{PI} = - \frac{\beta}{\alpha} \frac{\lambda}{Z^2} \Deriv{Z^2}{r}\Bigg|_\text{PI} ,
\end{align}
where $\Deriv{\cdot}{r} \big|_\text{PI}$ indicates the contribution to the evolution equation from just the pickup ion driving
  \citep{ZankEA96, MattEA99-swh}.

Recall, however, that  PI driving is expected to directly modify only the parallel length scale of the turbulence, 
and at scales of order the ion gyroradius, which is much smaller than the  turbulence correlation length 
     $ \lambda $
  \cite[e.g.,][]{LeeIp87, IsenbergEA03}.
Unfortunately, this physics is not accurately described in the 
  %% solar wind 
TTMs considered here, since these evolve a single characteristic lengthscale, $ \lambda $, that is both ``large''-scale and essentially a perpendicular rather than a parallel scale
\citep[see, e.g.,][]{OughtonEA11,Sokol.etal22, Zirnstein.etal22}.
Thus it is not a priori clear whether or not it is advantageous to include PI effects in the $\lambda$ equation,  as far as agreement with observational data is concerned.
  %% and we may investigate whether including PI forcing of this kind gives better or worse agreement with the observational data.

To investigate this we consider modified forms of the TTMs that \emph{exclude} the PI driving term in the $\lambda$ evolution equation, 
   %% (\autoref{eq:model-lambda}), 
which is thus replaced with
\begin{align}
    \label{eq:model-lambda-noPI}
      U \Deriv{ \lambda} {r}  &=
         \frac{\lambda \sigma_D M_\lambda U} {r} 
      +  \beta f^+(\sigma_c) Z .
\end{align}
We refer to these as ``no $\lambda$ PI'' TTMs, to be contrasted with the ``standard'' TTMs associated with equations~\eqref{eq:model-Z2}--\eqref{eq:model-temp}.
%% \ie{}we ignore the PI driving term in the $\lambda$ evolution equation. 
(Throughout this paper
 references to a specific analysis (e.g., $\alpha$--$\beta$--BC) 
 are to be interpreted as meaning that the TTM uses 
   \autoref{eq:model-lambda},
 unless we explicitly state that it instead uses
 \autoref{eq:model-lambda-noPI}.)

%  \green{In fact, 
%  the pickup ion driving occurs at roughly the ion gyroradius ($\rho_i$), close to the scales where energy dissipation starts to occur. 
%  Therefore, the $ \Epidot $ energy injection is occurring at small scales (i.e, not those modelled by the larger correlation length, $\lambda$). Moreover, pickup ion effects are expected to modify a characteristic length scale that is 
%   {parallel}
%  to the magnetic field, which our model does not describe. 
%  As part of our analysis we may thus investigate whether including such a PI forcing of $\lambda$ gives better or worse agreement with the observational data
%   (\autoref{sec:no_PI_analysis}).
% }

We may then
perform Bayesian analyses of the two situations, using the same setup as in \autoref{sec:ab_analysis} and \ref{sec:extended_analysis} and with priors as given in Tables \ref{tab:alpha_beta_prior} and \ref{tab:extended_priors}.
The results indicate that the ``no $\lambda$ PI'' TTM models 
 %% (i.e., using   \autoref{eq:model-lambda-noPI})
typically give worse agreement with the observational data than those employing
  \autoref{eq:model-lambda}.
    \autoref{tab:evidences}
lists the log-evidences for these cases and this will be discussed further in 
   \autoref{sec:discussion}.

  %% for model comparison, to determine whether the inclusion of the PI driving term in the $\lambda$ evolution equation is a useful modelling assumption to make. 
  %% Further discussion on this is deferred to 

  \section{Discussion}
  \label{sec:discussion}

%% Big picture/what can we learn?

  \subsection{Joint Posterior Distributions and their Correlations}
    \label{sec:joint}
   
%\YP{You can do this with getdist, see https://getdist.readthedocs.io/en/latest/chains.html#getdist.chains.WeightedSamples.cov}

%The Bayesian analysis provides posterior samples that can be used to calculate the joint-posterior distributions. \red{This describes the correlations of 2 or more variables, which can contain advantageous (or disadvantageous) information depending on the circumstances}. The vKH parameters, $\alpha$ and $\beta$, are elements of a closure model, and Bayesian inference assumes these parameters have distributions. This is in line with solar wind observations, that have determined the spectral powerlaw slope is not constant and has a distribution for separate observations \citep{BoldyrevEA11}, and observed distributions in the Reynolds number \citep{WrenchEA24}. As a closure model for the dissipation rate \citep[and therefore, the spectral cascade slope, and is related to the Reynolds number, see][]{MattEA96-jpp}, it might also be expected for $\alpha$ and $\beta$ to be distributed.

%\red{Need to reword this paragraph, we are attempting to provide justification for treating the constrained parameters as probability distributions according to Bayesian statistics.}

Typically, results from
solar wind TTMs have been presented using one, or a few, hand-picked values for the adjustable parameters. The Bayesian analysis techniques explored in this study improve on that situation by enabling study of the distributions of these parameters and illumination of their interrelations. 
The posterior samples obtained by the Bayesian analysis allow us to do so by analysing the joint-posterior distributions of the variables of interest. 
% \red{ \\ I'm not following what our point is in the below. \\ 
% "our" 5 params are not observational quantities. 
%  They're fitting constants (or more harshly fudge factors) to make a phenomenology or modelling term a quantitatively better match.  Whereas all the quantities in the para below are nothing to do with a theory, much less a phenomenology.  They're things solely and directly determined from analysis of observations.  So I'm not in favour claiming the two sets of variables are similar.\\
%  Ok to drop rest of this paragraph?\\ }
% \blue{MB: see comments.\\ }
% This puts the 
% \green{variables we constrain during the Bayesian analysis ($\alpha$, $\beta$, $\sigma_D$, $f_D$, $\Csh$ and the BCs)} 
% on a similar statistical footing 
% %\red{Not sure I u/stand here: what is it that's similar?}
% to observations of solar wind variables/parameters such as: the 
% slopes of magnetic and kinetic energy spectra \citep{BoldyrevEA11}, 
% fluctuation amplitudes and Mach numbers \citep{TuMarsch94, BavassanoBruno95}, 
% cross helicity \citep{TuMarsch95, BrunoCarbone13, BoldyrevEA13, AndresEA21}, 
% the inertial-kinetic range break-point \citep{ChenEA14,Pitna.etal21}, 
% proton temperatures \citep{HellingerEA13}, 
% the correlation scales \citep{CuestaEA22-intermit}, 
% and magnetic Reynolds number \citep{WrenchEA24}.

  \subsubsection[alpha--beta Correlation]{$\alpha$-$\beta$ Correlation}
    \label{sec:ab_correlation}

\begin{figure} % [hbt!]
\centering
\makebox[0cm]{\includegraphics[scale=1]{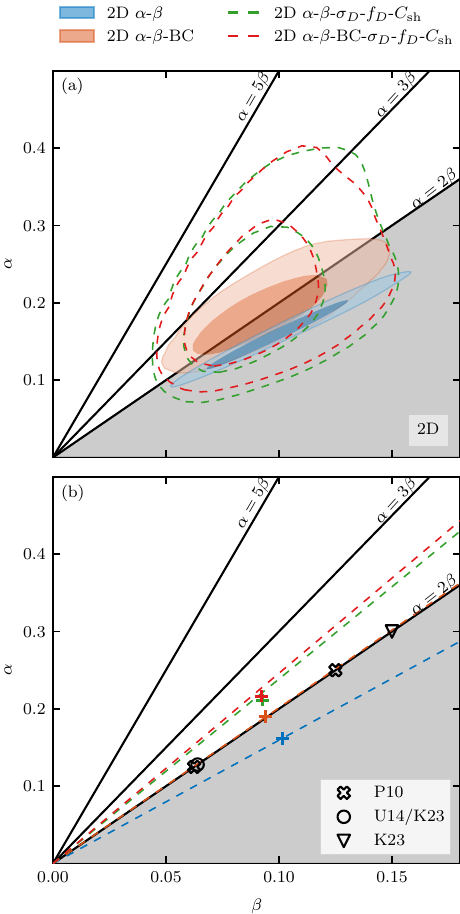}}
 \caption{(a) 
  $ 68\%$ and $95\%$ intervals of the joint $\alpha$--$\beta$ posterior distributions for the 2D TTM for several Bayesian analysis cases:
   $\alpha$-$\beta$ (blue contours), 
   $\alpha$-$\beta$-BC (orange contours), 
   $\alpha$-$\beta$-$\sigma_D$-$f_D$-$\Csh$ (green dashed), and 
   $\alpha$-$\beta$-BC-$\sigma_D$-$f_D$-$\Csh$ (red dashed). 
 (b) The corresponding constrained linear relations (dashed lines of the same colour) fitted to the weighted joint-posterior samples, with coloured crosses representing the means. 
 Some specific $\alpha$--$\beta$ pairs that have been used in the literature are also shown:
  P10 \citep{PeiEA10}, 
  U14 \citep{UsmanovEA14}, and 
  K23 \citep{KleimannEA23}.
 In both panels, the black lines represent 
  $ \alpha = m \beta$ relations, 
 corresponding to the physical implications associated with 
  constant Reynolds number ($m=2$), 
  Saffman's model ($m=3$), and 
  Kolmogorov's model ($m=5$). 
 The shaded region, $ \alpha \le 2\beta $,
 indicates values that imply the Reynolds number increases with heliocentric distance;
 these would be unphysical for purely decaying turbulence.
}
\label{fig:alpha_beta_relation}
\end{figure}

\hyperref[fig:alpha_beta_relation]{Figure 7a} shows the $\alpha$-$\beta$ joint posterior distributions for the 2D model analyses described in \autoref{sec:results}. 
Although the TTMs have forcing terms, 
one might wonder if these
 $\alpha$-$\beta$ correlations will nonetheless display behaviour related to the the conservation laws inherent in 
  von \karman{}--Howarth 
modelling for decaying homogeneous turbulence 
\citep{HossainEA95,MattEA96-jpp}, i.e., 
 $ \lambda Z^{2\beta/\alpha}  = const$. 
Particular values of
the ratio $m=\alpha/\beta$ correspond to some well-known turbulence models. For example, 
decay at constant Reynolds number ($m=2$), 
the \cite{Saffman67b} model ($m=3$), and 
the \cite{Kol41b} model ($m=5$),
which all yield straight lines in the $\alpha$--$\beta$ plane. 
To test this, we plot in \hyperref[fig:alpha_beta_relation]{Figure 7b} the constrained straight lines of best fit calculated using the weighted posterior samples. 
For comparison, we include specific pairs of $\alpha$--$\beta$ values that have been used in the literature 
  \citep{PeiEA10, UsmanovEA14, KleimannEA23}. 
We note that these literature values are not directly inter-comparable since the various studies typically use different TTMs (which could influence the posterior distributions of the 
 $\alpha$--$\beta$ parameters). 
Other studies often employed $ \alpha = 2 \beta$, but with $\beta \gtrsim 0.2 $ so that the values are off-scale in \autoref{fig:alpha_beta_relation}
 \citep[\eg{}][]{BreechEA08, AdhikariEA15, AdhikariEA17-NI-2, AdhikariEA17-NI-3, AdhikariEA21-xport}.

Clearly, the joint-distributions show strong positive correlations, particularly for the
 $\alpha$-$\beta$ and 
 $\alpha$-$\beta$-BC analyses. 
Indeed, the latter
  %% the $\alpha$-$\beta$-BC analysis 
(where uncertainty in the BCs is allowed for) 
strongly follows the $ \alpha = 2 \beta$ correlation. 
For homogeneous freely-decaying hydrodynamic turbulence this would be consistent with an approximately constant Reynolds number
  \citep[see][]{MattEA96-jpp}. 
However, the TTM models we analyze herein have shear and PI forcing and this alters expectations for the radial evolution of the Reynolds number \citep{ZankEA96}. 
Furthermore, observation-based estimates for (equivalent) Reynolds numbers for the solar wind have been shown to have considerable variability, both at $1\,\si{AU}$ and as a function of heliocentric distance 
  \citep{ParasharEA19, CuestaEA22-intermit, WrenchEA24}. 
Thus the reasons for an $\alpha = 2\beta$ correlation are likely to be more complicated than those holding in simpler decaying hydrodynamic situations.
 %\red{... drop next clause? but further analytical work is required \red{[SW16 proceedings]}}. 

Extending the analysis to include the parameters
 $\sigma_D$, $f_D$, and $\Csh$
yields 68\% intervals that lie largely in the region in between 
$\alpha=2\beta$ and $\alpha=3\beta$, 
with slightly steeper constrained linear relations
 ($\alpha \approx 2.38\beta$). 
Clearly, the interval contours are 
 significantly broader than those for the $\alpha$-$\beta$ and $\alpha$-$\beta$-BC analyses  and could be considered a superset of them. 
The $\alpha$-$\beta$ analysis has a shallower slope ($\alpha=1.59\beta$), and the entire distribution lies below the $\alpha = 2\beta$ line.
Recall that for freely-decaying homogeneous turbulence that line corresponds to a vKH phenomenology evolution where the Reynolds number is constant 
    \citep{Dryden43,MattEA96-jpp}. 
Moreover, $\alpha$--$\beta$ pairs lying below that line would correspond to dynamics with an increasing Reynolds number, which is unphysical for undriven turbulence.
However, in the TTMs we consider, driving is present and the $\alpha < 2\beta$ region is not necessarily forbidden and the overall picture can be significantly more complex
   \citep[cf.][Figure 1]{ParasharEA19}. 

  %%  \red{The weighted means of the posterior samples (the crosses in \autoref[b]{fig:alpha_beta_relation}) provide remarkably similar values for the different sets of analysis, \red{with small differences in the $\alpha$ value}.}

\begin{figure*} % [hbt!]
\centering
\makebox[0cm]{\includegraphics[scale=1]{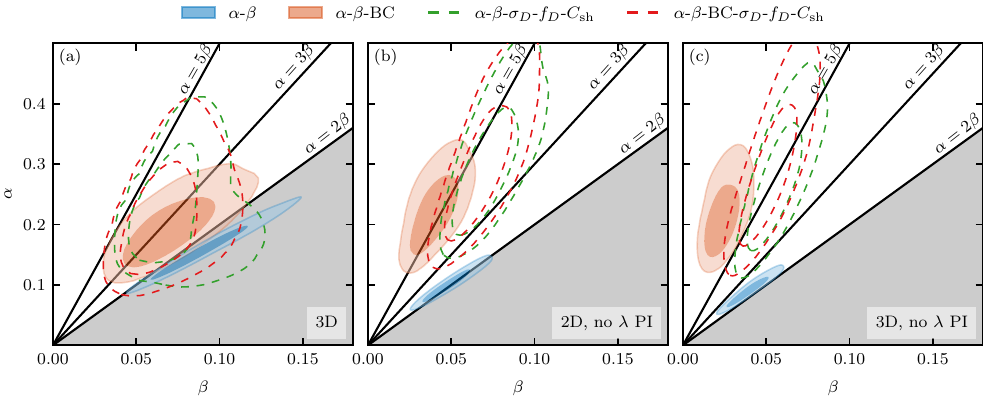}}
  \caption{As for
    \hyperref[fig:alpha_beta_relation]{Figure 7a} 
    %the joint   $\alpha$--$\beta$ posterior distributions for the
  except for the  
  (a) 3D TTM model, 
  (b) 2D TTM model with no PI term in the $\lambda$ equation, and 
  (c) 3D TTM model with no PI term in the $\lambda$ equation.
 Four analysis cases are shown in each panel: 
  $\alpha$-$\beta$    (blue contours), 
  $\alpha$-$\beta$-BC (orange contours), 
  $\alpha$-$\beta$-$\sigma_D$-$f_D$-$\Csh$ (green dashed), and 
  $\alpha$-$\beta$-BC-$\sigma_D$-$f_D$-$\Csh$ (red dashed).}
\label{fig:alpha_beta_relation_others}
\end{figure*}

\autoref{fig:alpha_beta_relation_others} is the equivalent of \hyperref[fig:alpha_beta_relation]{Figure 7a} but for some of the other analyses discussed in 
  \autoref{sec:results}.
The 3D model analysis $\alpha$-$\beta$ results 
(panel \hyperref[fig:alpha_beta_relation_others]{a}) 
share some similar behaviour, with constrained fitted linear correlations: $\alpha=1.75\beta$, $\alpha=2.70\beta$, $\alpha=3.33\beta$, $\alpha=3.33\beta$ for the $\alpha$-$\beta$, $\alpha$-$\beta$-BC, $\alpha$-$\beta$-$\sigma_D$-$f_D$-$\Csh$, and $\alpha$-$\beta$-BC-$\sigma_D$-$f_D$-$\Csh$ analyses respectively. The mean values for these sampled posterior distributions are centered around $\alpha \approx 0.21$, $\beta \approx 0.07$, with the exception of the 3D $\alpha$-$\beta$ analysis with its smaller mean $\alpha$ value of $0.16$. 
The contours of these analyses are consistent 
 (\ie overlapping to some degree) 
 and the posterior distributions from the $\alpha$-$\beta$-BC-$\sigma_D$-$f_D$-$\Csh$ analysis appear to be a superset of the smaller analyses.
  %\red{when we say consistent, are we meaning there is no reason to choose one vs another?}
    %\red{MB: this is noteworthy because (b) and (c) do not follow this behaviour.}

Panels (b) and (c) of 
\autoref{fig:alpha_beta_relation_others}
 %%  \hyperref[fig:alpha_beta_relation_others]{b} and \hyperref[fig:alpha_beta_relation_others]{c} 
display the 
$\alpha$--$\beta$ joint posteriors for the ``no $\lambda$ PI'' TTMs, respectively for the 2D model and the 3D model.
See \autoref{sec:no_PI_analysis}.
In both cases contours for the
$\alpha$-$\beta$ analysis strongly follow the $\alpha=2\beta$ line. 
The other 
  %% (excluded PI term in the $\lambda$ equation) 
analyses display steeper correlations and have contours distinct from the $\alpha$-$\beta$ analysis contours. 
It appears that the $\alpha$-$\beta$ analyses are too restricted, with the posterior constraints being driven to a different region of parameter space compared to the extended analyses of the same TTMs.  
This results in poor performance for these models, as discussed further in \autoref{sec:model_selection}.

   \subsubsection{Extended Correlations}
       \label{sec:other_correlations}

  %\Medit{There is a lot of information in these extended correlations, maybe we could write it up in a or something instead of this section?}
  %\Medit{The other option is to remove the correlation figures, and just have the joint-posterior figures. Mentioning we will dive into the correlations more rigorously in a later paper.}

\begin{figure*} % [hbt!]
\centering
\makebox[0cm]{\includegraphics[scale=1]{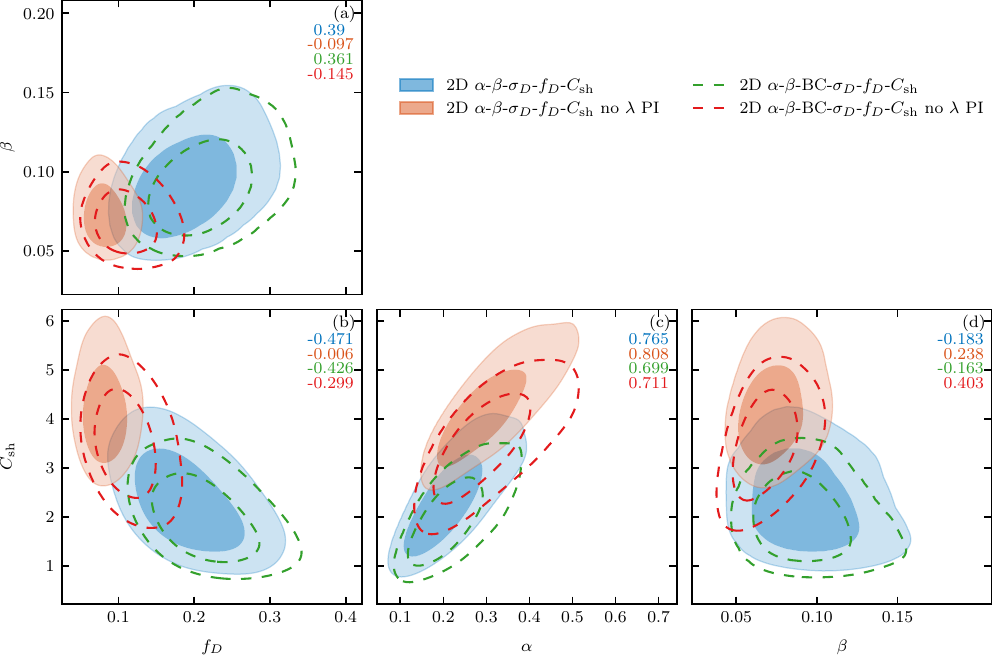}}
  \caption{Joint posterior distributions for the parameters $\alpha$, $\beta$, $f_D$, and $\Csh$ for four different 2D model analysis cases (see legend). 
    %% We show the $\alpha$-$\beta$-$\sigma_D$-$f_D$-$\Csh$, and $\alpha$-$\beta$-BC-$\sigma_D$-$f_D$-$\Csh$ 2D model analyses with (blue shaded contours, and green dashed contours, respectively) and without including 
  The two analyses labelled with ``no $\lambda$ PI'' 
%%    (filled beige and red dashed contours)
  refer to those discussed in \autoref{sec:no_PI_analysis}.
  Numerical values in the top right corners of each panel are the Pearson correlation coefficients.
  %\red{\\ Is it easy to move the legend down and across into the empty space in row 1?}
  }
\label{fig:alpha_driving_relation}
\end{figure*}
For the extended analyses (\autoref{sec:extended_analysis}) 
there are additional joint-posterior distributions to examine; i.e., 
not just those for $\alpha$ and $\beta$. 
  \autoref{fig:alpha_driving_relation} shows the joint posterior distributions for some of these, specifically for the
  parameters $\beta$-$f_D$, $\Csh$-$f_D$, $\Csh$-$\alpha$, and $\Csh$-$\beta$. 
We omit discussion of other joint-posterior correlations.
%%  Other joint-posterior distributions exhibit correlations, but for brevity, we choose not to discuss them.

%\begin{figure} % [hbt!]
%\centering
%\makebox[0cm]{\includegraphics[scale=1]{chapters/imgs4/small_corr_coefficients.pdf}}
%\caption{Correlations between sampled parameter posterior distributions. The $\alpha$-$\beta$-$\sigma_D$-$f_D$-$\Csh$ analyses with %(top right triangle) and without the $\lambda$ PI term (bottom left triangle).}
%\label{fig:small_corr_coefficients}
%\end{figure}
  %  \red{ add separate colour bars for different sections(?), change something}
  %  \red{ side-by-side triangles?}

%\begin{figure} % [hbt!]
%\centering
%\makebox[0cm]{\includegraphics[scale=1]{chapters/imgs4/large_corr_coefficients.pdf}}
%\caption{\Medit{Does this work properly for log-distributed parameters?}}
%\label{fig:large_corr_coefficients}
%\end{figure}

%\red{rewrite this paragraph}
%We show the 2D model $\alpha$-$\beta$-$\sigma_D$-$f_D$-$\Csh$ (shaded blue contours), and 2D model $\alpha$-$\beta$-BC-$\sigma_D$-$f_D$-$\Csh$ (green dashed line contours) analyses. We also show the same analyses where we have excluded the $\lambda$ PI term as shaded orange contours, and red dashed line contours respectively.

The `cleanest' set of correlations occurs for 
 $\Csh$-$\alpha$,
being  strong and positive for all models
(\hyperref[fig:alpha_driving_relation]{Figure 9c}).
%(\autoref{fig:alpha_driving_relation}c). 
Intuitively, when the energy injection due to shear increases, the dissipation rate will increase and, with vKH-style modeling, this can show up as an increase in the energy dissipation parameter $\alpha$. 
  %% This is in line with the Kolmogorov turbulence phenomenology \citep{Kol41a, Kol41b, Kol41c} \red{exact reference?}.

  % \red{ rewrite, the correlation present in the other analysis. There is no causality here.}

The two turbulence driving parameters, $\Csh$ and $f_D$, 
show a negative correlation for the `standard' 2D TTM
  (\hyperref[fig:alpha_driving_relation]{Figure 9b}),
but for the 2D TTM with no $\Epidot$ term in the $\lambda$ equation
  (\autoref{sec:no_PI_analysis})
 $\Csh$ is found to be almost independent of $f_D$. 
     %% with exception to the $\alpha$-$\beta$-$\sigma_D$-$f_D$-$\Csh$ analysis when the model has no $\lambda$ PI term. 
The negative correlation is an intriguing relation, as the shear driving term decays with radial distance ($\propto 1/r$), whereas $ \Epidot $ increases (nonlinearly)  with distance; 
  see~\autoref{eq:Epi-dot}.
The correlation may arise because 
as far as determining $\Csh$ and $f_D$ is concerned, the two forms of driving compete to some degree, and this is apparent in 
    %% their joint-posterior distribution correlations
  \hyperref[fig:alpha_driving_relation]{Figure 9b}.
For example, 
when the objective is for the TTM to match the observations,
making $\Csh$ larger is associated with a reduced requirement for PI energy injection (i.e., $f_D$ can be smaller).
Another possibility is that the correlation is not really physical, but rather an artifact connected with limitations of the modelling, such as $\Csh$ being constant at all distances, which conflicts with results from the split dataset analysis of \autoref{sec:split_analysis}. 
Further investigation is required, 
but these results indicate that the TTMs require energy injection to best match the observations, as is well known from previous studies
    \citep[e.g.,][]{ZankEA96,MattEA99-swh,RichardsonSmith03}.

 %% \red{Is our point (next) that the anticor is (or might be) an artifact of using $\Csh=const$, when probably it should change with distance?\\ }
 %% \red{However, this relation may be superficial due to the apparent change in $\Csh$ as the radial distance increases (\autoref{sec:split_analysis}).} 

TTMs that lack the $\Epidot$ term in the $\lambda$ equation show no, or weak, correlations of 
    $f_D$ and $\beta$
  (\hyperref[fig:alpha_driving_relation]{Figure 9a}).
This is expected since there is then no direct coupling of the PI driving to the $\lambda$ evolution; indirect effects, via $Z$, remain.
However, these ``no $\lambda$ PI'' TTMs show a positive correlation between 
$\Csh$  and $\beta$
  (\hyperref[fig:alpha_driving_relation]{Figure 9d}). 
Since $\Csh$ is not present in the $\lambda$ evolution equation (\autoref{eq:model-lambda}), this correlation must arise indirectly via couplings to the other equations. 
Conversely, 
for the ``standard'' TTMs
there is a weak positive correlation between $\beta$ and $f_D$ 
and little to no correlation between $\beta$ and  $\Csh$.

Analyzing the correlations and/or degeneracies of the posterior distributions gives hints at the physics being modelled 
 %% (unintentionally, or otherwise), 
and can help critique constraints on the model parameters. 
Some of these relations between pairs of parameters
are difficult to parse, in part due to the coupled nonlinear nature of the system of equations. 
% \red{However, their existence can be exposed, 
%    and we suggest that in some cases the correlations may be connected} 
%to known analytic conservation relations
%    \citep{ZankEA96, MattEA96-jpp, OughtonBishop-sw16}.

\subsection{Model Selection}
\label{sec:model_selection}

%\YP{Need to explain, either here or in the previous methodology section, that the evidence comparison protects against overfitting because it penalizes posteriors that cover large fractions of prior space.  So the fact that your fiducial model is favoured really means that there's not enough information in the data to constrain the extra parameters well.  I need to think about how the lack of errorbars affects this... EDIT: oh yes I see you explained this later.  I would move it up near the beginning (or into the methodology) to motivate why we can use the evidence for model selection.}

%\Medit{Try to explain why we think we can get away with doing the model comparison without error bars (they cancel).}

%\YP{I would calculate the log-evidence difference the other way around, so that a positive evidence difference in Table 5 means a model is favoured.  It's correct the way you've done it but just more usual to see it the other way around.}

To compare models we may use the Bayesian evidence values 
(\autoref{eqn-evidence})
to calculate the   \emph{Bayes factor}, 
  $ \ln K =   \ln \mathcal{Z}_{\text{model}} 
            - \ln \mathcal{Z}_{\text{fiducial}} $. 
This factor is often used in Bayesian analysis to help determine whether or not a model is more strongly supported by the data relative to a fiducial model. 
We use the commonly accepted Jeffreys' scale \citep{Jeffreys.Jeffreys98}: 
 $ 0.0 \le \abs{\ln K} < 1.0$ implies no          favourability, 
 $ 1.0 \le \abs{\ln K} < 3.0$ implies slight      favourability, 
 $ 3.0 \le \abs{\ln K} < 5.0$ is      significant favourability, and 
 $ 5.0 \le \abs{\ln K}$       is      decisive    favourability,
with the phrases no/slight/significant/decisive  favourability used as jargon. 
Since $\ln K = - \ln 1/K$, 
positive values imply reversed model favouring, \ie{}we favour the comparison model over the fiducial. 
Comparisons between two non-fiducial models can be quantified by subtracting the two Bayes factors. 
 %%  listed in \autoref{tab:evidences}.

% \pgfkeys{/pgf/number format/.cd, zerofill, precision=2}
% \newcommand{\STAB}[1]{\begin{tabular}{@{}c@{}}#1\end{tabular}}
% \begin{table}
% \centering
% \pgfplotstabletypeset[
%     col sep=comma,
%     %brackets/.style={
%     %    postproc cell content/.append style={/pgfplots/table/@cell content/.add={\relax(}{)}},
%     %},
%     columns={Analysis, 2D Model, 3D Model},
%     columns/Analysis/.style={string type, column type/.add={}{|}},
%     every head row/.style={before row=\hline\hline},
%     every row no 0/.style={before row=\hline\hline},
%     every row no 4/.style={before row=\hline},
%     %every row no 5/.style={before row=\rowcolor[gray]{.8}},
%     %every row no 6/.style={before row=\rowcolor[gray]{.8}},
%     %every row no 7/.style={before row=\rowcolor[gray]{.8}},
%     %columns/2D Model/.style={column type={r}},
%     %columns/3D Model/.style={column type={r}},
% ]{evidences_all.csv}
%   \caption{The Bayes factors, $\ln K = \ln \mathcal{Z}_{\text{model}} - \ln \mathcal{Z}_{\text{fiducial}}$, for the 2D and 3D models for the various analyses discussed in \autoref{sec:results}. We subtract the fiducial model log-evidence, chosen as that for the 2D $\alpha$-$\beta$ analysis, from each model log-evidence. 
%    The lower section of the table provides the Bayes factors for the TTMs \emph{without} the PI term in the $\lambda$ evolution equation (\autoref{sec:no_PI_analysis}), using the same 2D 
%     $\alpha$-$\beta$ 
%    fiducial model employed in the top section.}
% \label{tab:evidences}
% \end{table}

\begin{table}
\centering
\begin{tabular}{ c | c c }
    \hline\hline
    Analysis & 2D Model & 3D Model \\
    \hline\hline
     $\alpha$-$\beta$ &  0.00 & -5.40 \\
     $\alpha$-$\beta$-BC & -5.41 & -6.54 \\
     $\alpha$-$\beta$-$\sigma_D$-$f_D$-$\Csh$ & -4.35 & -9.57 \\
     $\alpha$-$\beta$-BC-$\sigma_D$-$f_D$-$\Csh$ & -10.61 & -11.48 \\
     \hline
     $\alpha$-$\beta$ & -32.97 & -42.11 \\
     $\alpha$-$\beta$-BC & -16.75 & -16.03 \\
     $\alpha$-$\beta$-$\sigma_D$-$f_D$-$\Csh$ & -9.70 & -11.97 \\
     $\alpha$-$\beta$-BC-$\sigma_D$-$f_D$-$\Csh$ & -15.44 & -16.30 \\
\end{tabular}
  \caption{The Bayes factors, $\ln K = \ln \mathcal{Z}_{\text{model}} - \ln \mathcal{Z}_{\text{fiducial}}$, for the 2D and 3D models for the various analyses discussed in \autoref{sec:results}. We subtract the fiducial model log-evidence, chosen as that for the 2D $\alpha$-$\beta$ analysis, from each model log-evidence. 
   The lower section of the table provides the Bayes factors for the TTMs \emph{without} the PI term in the $\lambda$ evolution equation (\autoref{sec:no_PI_analysis}), using the same 2D 
    $\alpha$-$\beta$ 
   fiducial model employed in the top section.}
\label{tab:evidences}
\end{table}

\autoref{tab:evidences} lists the Bayes factors obtained when the fiducial model is the 2D $\alpha$-$\beta$ analysis. The lower section is for analyses run using the model without the PI term in the $\lambda$ evolution equation (\autoref{sec:no_PI_analysis}). Bluntly, the fiducial model is significantly to decisively favoured relative to all the other analyses. 

For the analyses where the TTM's $\lambda$ equation does contain the 
$\Epidot$ term (top section of \autoref{tab:evidences}) the fiducial 2D $\alpha$-$\beta$ analysis is decisively favoured against the 3D $\alpha$-$\beta$ analysis. 
Similarly, the 2D $\alpha$-$\beta$-$\sigma_D$-$f_D$-$\Csh$ analysis is decisively favoured against its 3D version. 
These outcomes accord with
observational indications that MHD-scale solar wind fluctuations are predominantly quasi-2D in nature
  \citep[e.g.,][]{BelcherDavis71, MattEA90, BieberEA96, SmithEA06-aniso},
  and with 
theoretical and numerical expectations regarding anisotropy relative to a large-scale magnetic field 
   \citep[e.g.,][]{ShebalinEA83, CarboneVeltri90, OughtonEA94, MattEA96-var, GoldreichSridhar97, OughtonEA15, OughtonMatt20, Schekochihin-biased}.
The models with uncertainty in the BC conditions, 
  $\alpha$-$\beta$-BC, and $\alpha$-$\beta$-BC-$\sigma_D$-$f_D$-$\Csh$, 
have slight favourability to no favourability (respectively) for their 2D versions.

% By allowing uncertainty in the BCs, 
% \green{}
% \red{the boundary conditions get adjusted}
% to shift the curve to match the observations. 
% As shown by \autoref{fig:initial_distributions}, 
%   $\lambda_{\init}$ 
% is the only BC parameter to show a discrepancy between the 2D and 3D models. 

When we exclude the $\Epidot$ term from the $\lambda$ evolution equation (\autoref{sec:no_PI_analysis}), the 2D $\alpha$-$\beta$ analysis is still decisively favoured over the 3D. 
The 2D $\alpha$-$\beta$-$\sigma_D$-$f_D$-$\Csh$ analysis is only slightly favoured over its 3D model. 
The two analyses with the BCs show no favourability to either the 2D or the 3D model.

%\red{When $r \gg 1$: $\psi \rightarrow \pi/2$, $M_Z^{2D} \rightarrow 0$, and $M_{\lambda}^{2D} \rightarrow 1$. The 3D mixing terms are constant $M_{Z}^{3D} = M_{\lambda}^{3D} = 1/3$. words words words.} \blue{This was me starting to try to explain why the boundary conditions influence the favourability of the model so much.} 

By comparing the upper and lower sections of \autoref{tab:evidences}, we can assess whether including the PI driving term in the $\lambda$ evolution equation is a useful modelling assumption to make. 
We see that the fiducial model is favoured over any of the models that  exclude PI driving from the $\lambda$ equation. 
In fact, almost all combinations of the inclusion of the $\Epidot$ term in the $\lambda$ equation are decisively favoured compared to the same analysis excluding this term. 
Furthermore, the $\alpha$--$\beta$ analysis of TTMs without an $\Epidot$ term in the $\lambda$ equation is decisively disfavoured compared to any of the other analyses.
When the PIs are not directly influencing $\lambda$, the best result is decisively the 
$\alpha$-$\beta$-$\sigma_D$-$f_D$-$\Csh$ analysis, 
with slight favourability to the 2D model.

Our discussion in this section has shown the power of Bayesian analysis in connection with TTMs and we have seen that it can provide insights regarding model development. In particular, Bayesian analysis indicates that, of the models considered, the 2D TTM is a better fit than the 3D one, and that it is beneficial to include the PI driving term in the length scale equation. 
We also see a preference for the simpler analyses, that is, ones that constrain a smaller number of parameters. 
This is because the evidence is defined as the integral over the parameter space (\autoref{eqn-evidence}). 
Bayesian models with more parameters available will require 
  much lower $\chi^2$ (\autoref{eqn:chi2}), 
  relative to the simpler models,
  to compensate for the increase caused by integrating over the larger region. 
Visually, it appears that most models produce similar results 
  (compare \autoref{fig:ab_contour} and \autoref{fig:all_contour}), 
indicating that the Bayesian modelling of additional parameters does not net a significant accuracy boost. That is, the data apparently lacks sufficient information to constrain the extra parameters well. 

%In conclusion, the much simpler analysis, $\alpha$-$\beta$, assuming isotropy in 2D planes with respect to the large-scale magnetic field, and including the PI term in the $\lambda$ evolution equation is the favourite model (decisively).

%\red{What the choice of varying the initial condition means. Have assumed the steady-state of the equations, does it make sense for them to have a varying initial condition?}
%\textcolor{blue}{If we want a `best' solution, we can then use the MAP, or the weighted posterior mean.}
%\textcolor{blue}{Predicted functions based on the posterior mean of the parameters}
%\textcolor{blue}{We find the range of numbers the variable allows, given the model, that represents the data.}
%\red{We should be able to use this Bayesian analysis for physical insights into the choice of nonlinear and source terms and their relations, seen by the solar wind observations for choices of models.}
%Bayesian is getting the posterior of the parameters.

    \section{Conclusions}
      \label{sec:conclusion}

%\YP{I personally like shorter, snappy conclusions.  I usually number mine which helps to focus down to the essentials.  This is a personal preference though and you don't have to do it if you don't like it.}

We have shown that Bayesian analysis can be used to constrain values of adjustable parameters present in a class of energy-containing solar wind turbulence transport models.
%The analysis is able to 
%  \red{retrieve ??} 
%known physics, and also probe unknown physics. 
The analysis also provides quantification regarding the consistency of a TTM with observational data.
Additionally, we have shown how Bayesian evidence (in the form of readily-computed Bayes factors) may be used to objectively compare distinct TTMs and thus select a preferred TTM.
Our solar wind data comes from two sources: the \emph{Parker Solar Probe} and \emph{Voyager}~2 spacecraft missions. We have analyzed the compatibility of the solar wind datasets by performing the Bayesian analysis on the two datasets separately, and then together. The analysis with the combined datasets is able to place stronger, and consistent (with the separate dataset analyses) constraints on the solar wind transport model parameters. 
We also analyzed how the TTM parameters might vary between conceptually distinct physical regions, such as inside and outside 
    5\,\si{AU}.
The results match physical expectations that the inner heliosphere data predominates for constraints on the initial conditions and $\Csh$ parameter, while the outer heliosphere data better constrains the strength of the PI driving, $f_D$. 
We find a slight discrepancy between the inner and outer heliosphere values for $\Csh$, suggesting 
that this approach to modelling large-scale velocity shear 
  (i.e., $ \Csh U Z^2/r $) has shortcomings. 
\response{We note that other models for the shear driving exist
  \citep[e.g.,][]{WiengartenEA15, ZankEA17-NIxport}. 
  In particular, \cite{ZankEA17-NIxport} propose a model with $r^{-2}$ dependence 
  %% (rather than our $ Z^2(r)/r $) 
and this different scaling may be able to better address the decrease in the $\Csh$ parameter with radial distance associated with the present Bayesian analysis. Investigation of this is left for future work.}

%\blue{recap modelling correlations.}\\
The joint-posterior distributions display behaviour consistent with known behaviour of the von \karman{}--Howarth parameters for homogeneous turbulence. We have also investigated the link between the von \karman{}--Howarth parameters with the turbulence driving coefficients. These relations imply extended forms of the homogeneous turbulence conservation property $Z^{2\beta/\alpha} \lambda = \mathrm{const}$.

Using the Bayesian evidences, we decisively rule out the 3D isotropic turbulence assumption, in favour of fluctuations that are isotropic in 2D planes with respect to the large-scale magnetic field. This is consistent with observed and expected (anisotropic) dynamics in the solar wind. The model evidences also indicate that explicit inclusion of pickup ion effects in the correlation length equation---with
 the term obtained by assuming that
   $ Z^{2\beta/\alpha} \lambda $ 
 is locally conserved---is 
decisively favoured over excluding this term.

\response{The analysis discussed in \autoref{appendix:aT_analysis} indicates that,
for the considered TTMs,
the fraction of cascaded turbulence energy that heats solar wind protons is $0.2 \lesssim \alpha_{T} \lesssim 1$ 
and that this depends nonlinearly on the $\alpha$, $f_{D}$, and $\Csh$ values. 
Given the small nominal estimates for $f_{D}$ and $\Csh$ (\autoref{tab:common_model_parameters}),
we expect $0.6 \lesssim \alpha_{T} \lesssim 1$. 
Moreover, although we use $\alpha_{T} = 1$ in the main body of the paper, the considered TTMs are not very sensitive to the specific value of $\alpha_{T}$ used. 
Indeed, TTM solutions using $\alpha_T$ in this range have radial profiles that are visually very similar 
(see, e.g., \autoref{fig:all_contour} and \autoref{fig:Talph_contour}).
We intend to investigate this more thoroughly in future work.}

Consequently, based on the analysis of the particular TTMs and datasets employed herein, we recommend use of the 2D TTM with 
    %$\alpha \approx 0.16$, 
    $\alpha = 0.16 \pm 0.03$, 
    %$\beta \approx 0.10$
    $\beta = 0.10 \pm 0.02$,
    the parameters stated in \autoref{tab:common_model_parameters},
and
 inclusion of PI effects in the lengthscale evolution equation.
Naturally, these parameter values 
(as determined by the posterior distribution \autoref{eqn:bayes_theorem})
depend on both the particular datasets analysed, and the class of TTMs. Nonetheless, given the considerable scatter of much of the observational data we anticipate that the values may be relatively robust for the solar wind conditions considered.

%\blue{recap model selection.}\\
%Using the Bayesian evidences, we decisively rule out the 3D isotropic turbulence assumption, in favour of isotropy in 2D planes with respect to the large-scale magnetic field. This is consistent with expected dynamics in the solar wind. Additionally, we investigate ambiguity in the modelling through the pickup ion behaviour in our single correlation length. We have determined that the inclusion of a pickup ion term in the correlation length equation, derived by conserving the integral scale, is decisively favoured.

%future work would involve taking into account:
%better and more data
%more complicated models (see kleimann 6 equation model, and adhikari/zank 6 eqn model),
%errors
%physics, can look at different regions and compare, i.e. before pickup ions, can discuss differences in nonlinear models, nonlinear parameter space (alpha-beta relation), different datasets etc.

In closing, we note that the Bayesian analysis may be readily extended to more sophisticated solar wind models and their parameters.
Such extensions might involve the six equation models of
\cite{AdhikariEA15, AdhikariEA21-xport}, \cite{ShiotaEA17}, and \cite{KleimannEA23},
\response{
inclusion of a decelerating solar wind
  \citep{WangRichardson03, IsenbergEA10, ElliottEA19},
  two-component models
    \citep{OughtonEA11,WiengartenEA16,ZankEA17-NIxport},
    and
  dynamic large-scale fields
  \citep{UsmanovEA11,UsmanovEA16, WiengartenEA15,KleimannEA23}.}
%  
% such as the six equation models of \cite{AdhikariEA15, AdhikariEA21-xport} and \cite{KleimannEA23}.
% %\response{, including terms of order $V_{A}/U$, as well as changes in the mean solar wind speed $U$ \citep{IsenbergEA10, IsenbergEA-sw11, ZankEA18-pui, ElliottEA19}}. 
% \response{We have also neglected that, due to the pickup ions, $U$ should decrease to a value that is $\approx 10\%$ slower at $r \sim 60$\,\si{AU} than at 1\,\si{AU} 
%   \citep{WangRichardson03, ElliottEA19} 
% which could result in increased $T$ at large radial distances \citep{IsenbergEA10}. This effect could change the parameter constraints obtained in this paper, although we expect the change to be small.} 
%
The analysis may also be employed with space weather models and might lead to improved predictions of, for example, solar flare and CME interactions with earth.

%\Medit{Probably worth mentioning, it could be interesting to add in intrinsic noise/variability to the model equations. Then solve for this term (using the nested sampling) to give an idea of the variability. ATM there is no way of distinguishing between the variability in the solar wind, vs error in the data.}

\section*{Author Contributions}
MAB: conceptualization, analysis, interpretation, discussions, manuscript preparation \& writing.\\
SO: transport model expertise, interpretation, discussions, significant manuscript writing \& review.\\
TNP: interpretation, discussions, manuscript writing \& review.\\
YCP: Bayesian analysis expertise, interpretation, discussions, manuscript review \& editing.
%\begin{itemize}
%    \item MAB: conceptualization, data analysis, manuscript preparation
%    \item SO: solar wind transport model expertise, interpretation, discussions, manuscript editing
%    \item TNP: interpretation, discussions, manuscript editing
%    \item YCP: Bayesian analysis expertise, interpretation, manuscript review
%\end{itemize}

\section*{Acknowledgements}
%\begin{acknowledgments}
%% MAB and YCP are 
Research supported by the Marsden Fund Council from  NZ Government funding, managed by Royal Society Te Apārangi (E4200). We are grateful to
Dr C.\ W.\ Smith  for kindly providing access to the proton temperature \emph{Voyager} 2 solar wind observational dataset from their paper
\citep{SmithEA06-pi}. 
We also thank Dr L.\ Adhikari for providing the derived turbulence energy and correlation length, and proton and electron temperature data from his papers \citep{AdhikariEA15, AdhikariEA21-xport}. These data were derived from NASA Parker Solar Probe, Voyager 2, and Ulysses magnetometer and plasma data sets. \response{We thank the anonymous referee for  helpful suggestions that have improved the presentation of this paper.}
  %% for kindly providing the observationally derived turbulence energy and correlation length, and proton temperature data from his paper \citep{AdhikariEA21-xport}. These data were derived from NASA \emph{Parker Solar Probe}, \emph{Voyager 2}, and \emph{Ulysses} magnetometer and plasma data sets. The NASA \emph{Voyager} 2 turbulence energy, and correlation length data were also provided by Dr Laxman Adhikari, from his paper \cite{AdhikariEA15}.The NASA \emph{Voyager 2} proton temperature data was obtained courtesy of \cite{SmithEA06-pi}.
%\end{acknowledgments}

%% TODO: Put footnotes when mentioning software in the paper to the homepage of the code.
%% Add in the references

\software{Python:
    multinest,
    pymultinest,
    fgivenx,
    getdist,
    sympy,
    matplotlib,
    astropy,
    numpy,
    scipy,
    pandas.}

\appendix

%--------------------------------------------
\clearpage
    \section{Updated Breech et al.\ 2008 Model}
      \label{appendix:breech_model}

\cite{BreechEA08} presented a four-equation model for the time-steady radial transport of solar wind fluctuations in prescribed background solar wind fields:
\begin{align}
  \label{eq:Breech-Z2}
   U \Deriv{Z^2}{r}  &= 
    - \frac{UZ^2}{r} \left[
           1 +  \sigma_D M_Z - \Csh
       \right]
    + \Epidot
    - \alpha f^+(\sigma_c) \frac{Z^3}{\lambda } ,
 \\
  \label{eq:Breech-lambda}
   U \Deriv{ \lambda} {r} &= 
        \beta f^+(\sigma_c) Z  
      - \frac{\beta}{\alpha} \frac{\lambda}{Z^2} \Epidot ,
\\
  \label{eq:Breech-sigc}
   U \Deriv{\sigma_c} {r}  &= 
     \alpha f'(\sigma_c) \frac{Z }{ \lambda}
     -\left[\frac{U}{r} \left( \Csh - \sigma_D M_Z \right)
        +  \frac{\Epidot} {Z^2}
      \right] \sigma_c ,
 \\
  \label{eq:Breech-temp}
   U \Deriv{T} {r} &= 
      -\frac{4}{3}\frac{TU}{r}
      + \frac{\alpha}{3}\frac{m_p}{k_B}
         f^+(\sigma_c) \frac{Z^3}{\lambda} ,
\end{align}
where we employ the same notation used in the body of the present paper.

This is actually two models---one 
 for fluctuations that are (3D) isotropic 
and 
 one for fluctuations that are 2D with respect to the large-scale magnetic field (and isotropic in the 2D planes).  
The distinction between the models enters via the (energy) mixing operator 
\begin{equation}
   M_Z = 
   \begin{cases}
      \frac{1}{3}, & \text{ 3D} \\
      \cos^2 \psi, & \text{ 2D}
   \end{cases}
 \label{eq:Mz}
 .
\end{equation}

Unfortunately a mistake was made during the derivation of the lengthscale equation~\eqref{eq:Breech-lambda}.
Specifically, when integrating their equation (17),
over the lag $ {\zeta} $, the $\zeta=0$ forms of 
    $R_{ij}^{D,\text{s}}(\vec{\zeta}) $
were used rather than the full correlation functions
  \citep[cf.\  Appendix B in][]{KleimannEA23}.  
Rectifying this yields the updated equation for $L(r) = \int_0^\infty R_{jj}(r, \vct{\zeta}) \,\d\zeta $
and thence the updated equation for $\lambda = L/Z^2$:
\begin{align}
%%   \Deriv{L} {r} & = ...    
%%  , \label{eq:L-corrected}  \\  
  U \Deriv{ \lambda} {r}  &=
         \frac{\lambda \sigma_D M_\lambda U} {r} 
      +  \beta f^+(\sigma_c) Z
      -  \frac{\beta}{\alpha} \frac{\lambda}{Z^2} \Epidot,
  \label{eq:-lambda-corrected}
\end{align}
which is \autoref{eq:model-lambda} in \autoref{sec:TTM}.
This contains a new term involving the $\lambda$ mixing operators
\begin{equation}
   M_\lambda = 
   \begin{cases}
      \frac{1}{3}, & \text{ 3D} \\
      \sin^2 \psi, & \text{ 2D}
   \end{cases}
 \label{eq:Mlambda}
 ,
\end{equation}
where the closure $\lambda_D = \lambda$ has been used in $L_D = \lambda_D D $.

Note that for the model evolving 2D fluctuations there is a subtlety to the interpretation of $\lambda$.  To make contact with solar wind observations, correlation functions are calculated using \emph{radial} lags---rather than lags in the 2D plane. 
Consequently, if
  $ \lambda^{2D} $
is the in-plane correlation length, the $\lambda$ appearing in the model equations above (obtained after integrating the correlation functions over radial lag) is actually $\lambda^{2D} / \sin \psi $,
which is larger than the `true' value.

%% \cite{BreechEA08,KleimannEA23}

%%\subsection{Isotropic}

%\begin{table}
%\centering
%\begin{tabular}{c c}
%\hline
%\hline
%Parameters & Values\\
%\hline
%% $M$ & $1/3$\\
%$U$ & $400$ \si{km/s}\\
%$\sigma_D$ & $-1/3$\\
%$r_A = \frac{1 + \sigma_D}{1 - \sigma_D} $ & $1/2$  (\text{not independent so only need one of $r_A$, %$\sigma_D$})\\
%\hline
%\end{tabular}
%\caption{Is there a reason this is a separate table, rather than combined with the previous table?}
%\label{results:tab:alpha_beta_parameters}
%\end{table}

%% \subsection{Anisotropic}
    \section{Turbulence Dissipation and Proton and Electron Heating}
        \label{appendix:aT_analysis}
\response{In the solar wind, not all of the cascaded turbulence energy heats solar wind protons, with some of it instead heating electrons  \citep[see,  \eg{}][]{BreechEA09, CranmerEA09, Howes11, AdhikariEA21-xport, BandyopadhyayEA23-peHeat}. 
The details governing this partitioning are not well understood and thus represent another interesting candidate to apply Bayesian analysis to.}

\response{To investigate this 
we modify the turbulent heating term in the (proton) temperature \autoref{eq:model-temp} to include an additional parameter  $\alpha_{T} \in [0, 1]$,}
\begin{align}
    \label{eq:model-temp-alphT}
    U \Deriv{T} {r} &= 
      -\frac{4}{3}\frac{TU}{r}
      + \alpha_{T} \frac{\alpha}{3}\frac{m_p}{k_B}
         f^+(\sigma_c) \frac{Z^3}{\lambda} .
\end{align}
\response{This parameter describes the fraction of turbulence energy that is injected into the proton temperature (with the rest assumed to go towards electron scales). The Bayesian analyses in \autoref{sec:results} can be thought of as having a $\delta$-function prior 
 %%(\ie{}kept fixed) at 
 that mandates $\alpha_{T} = 1$. 
We perform two 2D TTM Bayesian analyses: $\alpha$-$\beta$-$\alpha_{T}$, and $\alpha$-$\beta$-BC-$\sigma_{D}$-$f_{D}$-$\Csh$-$\alpha_{T}$ using the corresponding priors stated in \autoref{tab:alpha_beta_prior} and \autoref{tab:extended_priors}, 
and using a uniform distribution on $ [0.1,1] $ 
as the prior for $\alpha_T$.}

%%with \autoref{tab:alphaT_prior}.}

% \begin{table}[!htb]
% \centering
% \begin{tabular}{c c}
% \hline
% \hline
% Parameter & Prior distribution \\
% \hline
% \hline
% $\alpha_{T}$ & $\mathcal{U}[0.1, 1]$\\
% \hline
% \end{tabular}
% \caption{\response{Prior distribution of the $\alpha_{T}$ parameter used in the Bayesian analysis of the 2D TTM. $\mathcal{U}[\cdot, \cdot]$ denotes a uniform distribution with the arguments indicating the domain boundaries.}}
% \label{tab:alphaT_prior}
% \end{table}

\begin{figure*} % [hbt!]
\centering
\makebox[0cm]{\includegraphics[scale=1]{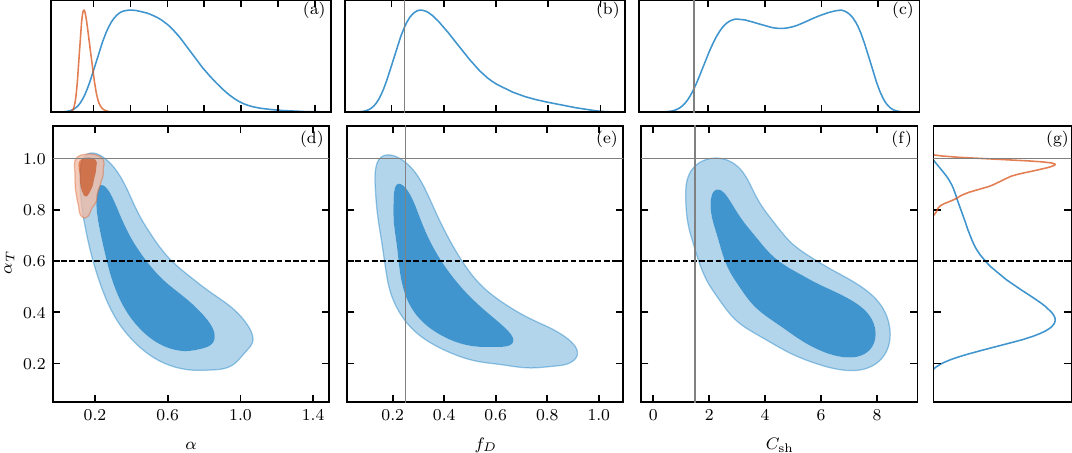}}
\caption{\response{2D $\alpha$-$\beta$-$\alpha_{T}$ (orange) and 2D $\alpha$-$\beta$-BC-$\sigma_{D}$-$f_{D}$-$\Csh$-$\alpha_{T}$ (blue) joint posterior distributions for the parameters $\alpha_{T}$ with $\alpha$, $f_{D}$, and $\Csh$ 
(panels d, e, and f, respectively) 
as well as 1D posterior distributions for $\alpha$, $f_{D}$, $\Csh$, and $\alpha_{T}$ 
(panels a, b, c, and g, respectively). 
The solid gray vertical and horizontal lines are the 
    \response{nominal} estimates, 
\ie{}the assumed values for the analyses that do not constrain these variables; see \autoref{tab:common_model_parameters}. 
The horizontal black dashed line at $\alpha_{T} = 0.6$ is as reported on in 
\cite{BreechEA09} and  \cite{CranmerEA09}.}}
\label{fig:degeneracies_Talph}
\end{figure*}

\response{\autoref{fig:degeneracies_Talph} displays some of the joint posterior distributions for the 2D $\alpha$-$\beta$-$\alpha_{T}$ and $\alpha$-$\beta$-BC-$\sigma_D$-$f_{D}$-$\Csh$-$\alpha_{T}$ analyses. 
Note the ``banana'' shape of the $\alpha$-$\beta$-BC-$\sigma_D$-$f_{D}$-$\Csh$-$\alpha_{T}$ posterior distributions.  Such shapes can be problematic to interpret and indicate that the associated 1D posteriors can be misleading.
  %%  We first warn the readers, when posterior distributions of this shape are found, you must examine the multi-dimensional posteriors rather than the 1D posteriors. 
For example, the 1D posterior distribution for $\alpha_{T}$ (\hyperref[fig:degeneracies_Talph]{Figure 10g}) seems strongly peaked at $\approx 0.4$. However, the 95\% confidence intervals for the joint-posterior distributions for $\alpha_{T}$ indicate a (much wider) range between 0.2--0.9. 
We also see that
the 2D $\alpha$-$\beta$-$\alpha_{T}$ analysis is strongly pushed up against the upper boundary of the $\alpha_{T} $ prior (with $\alpha_T > 1 $ being unphysical behaviour) and gives mean values of $\alpha_{T} \approx 0.93$ and an $\alpha$ and $\beta$ that are equal to those from the 2D $\alpha$-$\beta$ analysis.} 
\response{The joint posteriors thus indicate that the choice of the $\delta$-function prior for $\alpha_{T} = 1$ effectively restricts the posterior distributions of the other parameters ($\alpha$, $f_{D}$, and $\Csh$) to the smaller values obtained in \autoref{sec:results} and \autoref{sec:discussion}.}

\begin{figure*} % [hbt!]
\centering
\makebox[0cm]{\includegraphics[scale=1]{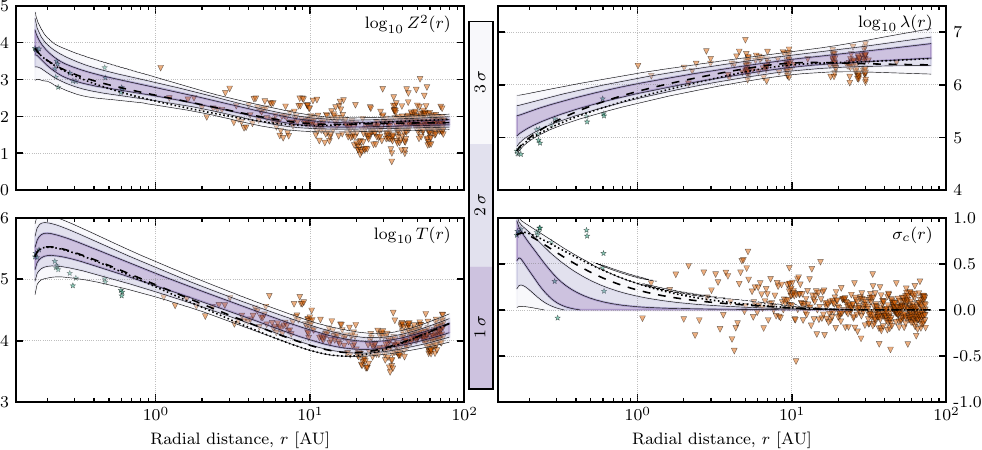}}
\caption{\response{Solutions for the 2D $\alpha$-$\beta$-BC-$\sigma_{D}$-$f_{D}$-$\Csh$-$\alpha_{T}$ analysis. The contours represent the confidence intervals of the means in the solutions obtained by the sampled posterior distributions. 
Green stars indicate \emph{PSP} data 
and orange triangles \emph{Voyager} 2 data. 
The 2D and 3D TTM posterior sampled means of the $\alpha$-$\beta$ analysis (\ie{}with $\alpha_{T} = 1$) are plotted with dashed and dotted black lines respectively. 
Units are as in \autoref{tab:observation_initial_values}.}}
\label{fig:Talph_contour}
\end{figure*}

\response{ 
Solutions to the TTM (with $\alpha_T$) are shown in 
    \autoref{fig:Talph_contour}.
Comparing this with
     \autoref{fig:all_contour}
one sees that the introduction of $\alpha_T$ produces only small differences.
  %% shows there is minimal difference in the distribution of solutions for the 2D $\alpha$-$\beta$-BC-$\sigma_{D}$-$f_{D}$-$\Csh$-$\alpha_{T}$ analysis (compare to \autoref{fig:all_contour}).
This similarity suggests that the TTMs contain more parameters than can be constrained effectively by the available observational data 
(see \autoref{sec:model_selection}).}

\response{Log-evidences for the two $\alpha_{T}$ analyses are given in \autoref{tab:evidences-alphaT} and indicate only slight favourability for the 2D TTM $\alpha$-$\beta$ analysis over the 2D TTM $\alpha$-$\beta$-$\alpha_{T}$ analysis. There is no favourability for the 2D $\alpha$-$\beta$-BC-$\sigma_{D}$-$f_{D}$-$\Csh$ analysis over the 2D $\alpha$-$\beta$-BC-$\sigma_{D}$-$f_{D}$-$\Csh$-$\alpha_{T}$ analysis. In other words, adding the $\alpha_{T}$ parameter has not yielded any better performance of the TTM (this is supported by the similar solutions in \autoref{fig:Talph_contour} to \autoref{fig:all_contour}). The simplest 2D TTM $\alpha$-$\beta$ analysis remains the favoured model, likely for similar reasons to those discussed in \autoref{sec:model_selection}.}

% \response{Log-evidences for the two $\alpha_T$ analyses are given in \autoref{tab:evidences-alphaT} and indicate only slight favourability for the 2D $\alpha$-$\beta$ analysis over the 2D $\alpha$-$\beta$-$\alpha_{T}$ analysis, but decisive favourability over the 2D $\alpha$-$\beta$-BC-$\sigma_{D}$-$f_{D}$-$\Csh$-$\alpha_{T}$ analysis. 
% Similarly, there is no favourability for the 2D $\alpha$-$\beta$-BC-$\sigma_{D}$-$f_{D}$-$\Csh$ analysis over the 2D $\alpha$-$\beta$-BC-$\sigma_{D}$-$f_{D}$-$\Csh$-$\alpha_{T}$ analysis. 
% In other words, adding the $\alpha_{T}$ parameter has not yielded any better performance of the TTM. 
% The simplest 2D $\alpha$-$\beta$ analysis remains the significantly favoured model, likely for similar reasons to those discussed in \autoref{sec:model_selection}, 
% .... \\
% and the previously chosen $\alpha_{T} = 1$ value is consistent with the data, given the TTM.
%  \\ \vdots  ... SAY MORE ... \\
%   ... like this suggests the TTM has shortcomings since observations are pretty clear that $\alpha_T $ is often significantly less than 1...
% }

\response{Considering the shape of the $\alpha$-$\beta$-BC-$\sigma_{D}$-$f_{D}$-$\Csh$-$\alpha_{T}$ joint-posterior distributions in \autoref{fig:degeneracies_Talph}, with small estimates for $f_{D}$ and $\Csh$, we see similar preference in the Bayesian analysis for values 
 $ 0.6 \lesssim \alpha_{T} \lesssim 1$. 
Imposing 
the \response{nominal} estimates for $f_{D}$ and $\Csh$ (\autoref{tab:common_model_parameters}) leads to a reduced range of $0.8 \lesssim \alpha_{T} \lesssim 1 $.}

\response{Given the observational results that
  $\alpha_{T} \sim 0.6$ \citep{BreechEA09, Cranmer09},
this set of features---the 
range of likely $\alpha_{T}$ values, 
the log-evidence values in \autoref{tab:evidences-alphaT}, 
and 
the similarity of the solutions shown in \autoref{fig:all_contour} and \autoref{fig:Talph_contour}---suggests that the TTMs considered herein may have shortcomings, as far as matching observational data is concerned.  
An alternative possibility is that more observational data is required to adequately constrain the TTM parameters.}

\begin{table}
\centering
\begin{tabular}{ c | c }
 \hline\hline
 Analysis & 2D \\
 \hline\hline
 $\alpha$-$\beta$-$\alpha_{T}$ & -1.87 \\  
 $\alpha$-$\beta$-BC-$\sigma_{D}$-$f_{D}$-$\Csh$-$\alpha_{T}$ & -10.28
\end{tabular}
\caption{\response{Bayes factors, $\ln K = \ln \mathcal{Z}_{\text{model}} - \ln \mathcal{Z}_{\text{fiducial}}$, for the 2D TTMs  that use \autoref{eq:model-temp-alphT}. 
To obtain $ \mathcal{Z}_{\text{fiducial}} $ we use the 2D $\alpha$-$\beta$ analysis (see \autoref{tab:evidences})}.}
\label{tab:evidences-alphaT}
\end{table}
%========================================

    \section{Bayesian Methodologies}
      \label{appendix:mcmc}

Evaluation of the evidence $\mathcal{Z}(\mathcal{D})$ (\autoref{eqn-evidence}) is usually ignored for parameter estimation since it is independent of the parameters $\Theta$. With this approach, parameter estimation inferences are typically obtained by taking samples from the unnormalized posterior, for example using Markov chain Monte Carlo (MCMC) methods. However, for  comparison of models $\mathcal{Z}(\mathcal{D})$ is required and evaluation of this multidimensional integral can be numerically challenging. 

%Evaluation of this multidimensional integral can be numerically challenging, so, for parameter estimation, it is usually ignored since it is independent of the parameters $\Theta$. With this approach parameter estimation inferences, like Markov chain Monte Carlo (MCMC) methods, are obtained by taking samples from the unnormalized posterior.

To calculate $\mathcal{Z}(\mathcal{D})$, we use \emph{nested sampling} \citep[see,][for a review on nested sampling, and its use with physical applications]{Ashton.etal22}, which reformulates $\mathcal{Z}(\mathcal{D})$ 
  %% (\autoref{eqn-evidence}) 
into a one-dimensional integral over contour regions of the likelihood \citep{Skilling04, Feroz.Hobson08, Feroz.etal09}:
\begin{align}
    \mathcal{Z}(\mathcal{D}) = \int_0^1 \mathcal{L}(X)\,\d X ,
\end{align}
where
\begin{align}
    X(\lambda) = \int_{\mathcal{L}(\Theta)>\lambda} \pi(\Theta)\,\d^D \Theta .
\end{align}
This simplifies the calculation, and provides an efficient method for providing parameter constraints and model evidences. By \textit{independently} selecting parameter values $\Theta$, we can evaluate the corresponding likelihood, that is,  how likely it is that these chosen parameters are able to reproduce the observed data. Recursively sampling new parameter values allows iterative estimation of their (posterior) distributions. These may be employed to constrain parameters present in solar wind transport models, and the evidence $\mathcal{Z}(\mathcal{D})$ used to compare different models.

\begin{figure}
\centering
\makebox[0cm]{\includegraphics[scale=1]{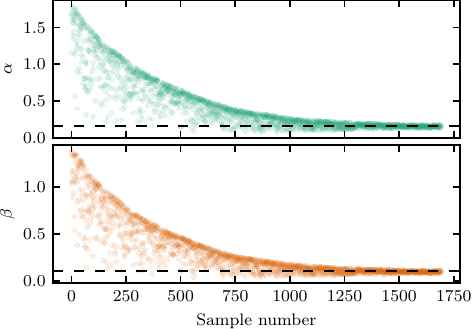}}
\caption{The samples of the 2D $\alpha$-$\beta$ analysis, showcasing the convergence of the $\alpha$-$\beta$ parameters towards the mean sampled posterior values (horizontal, black, dashed line).}
\label{fig:convergence}
\end{figure}

In our analysis we use the python interface
\texttt{pymultinest} \citep{Buchner16}
to \texttt{multinest} \citep{Feroz.etal09}, where the latter implements a multimodal nested sampling algorithm. This Bayesian inference tool calculates the evidence and produces posterior samples from distributions that may be multimodal, or exhibit complex degeneracies in high dimensional data. Using \texttt{pymultinest}, we can obtain constraints on the parameters by repeatedly solving the model with different parameter values. 
Approximate analysis runtimes, and the number of likelihood evaluations (roughly the number of times we need to solve the system of differential equations) are provided in \autoref{mcmc:tab:time}. 
By default, \texttt{multinest} automatically runs until the parameter sampling has converged; this is essential to obtain accurate 
  %% for the 
integration over likelihood contour levels. We use the number of live points as $100$ times the number of parameters we are sampling, with a default sampling efficiency of 0.8. An example of convergence of parameter sampling is shown in \autoref{fig:convergence}, where the $\alpha$ and $\beta$ parameters are plotted as a function of their sampling number (\ie{}simply plotting the chains provided by \texttt{multinest}) which corresponds to the time of sampling. In the case of both parameters, the whole prior distribution is sampled to start with, then the chains converge towards the mean sampled posterior value. The output of \texttt{multinest} can provide the weighted mean, maximum likelihood, or maximum a posteriori estimate. This output is used to determine a `best' solution for the constrained parameters.
 
\begin{figure}
\centering
\makebox[0cm]{\includegraphics[scale=1]{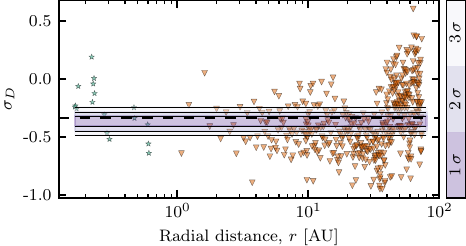}}
\caption{Same format as \autoref{fig:all_contour}. Solutions for $\sigma_D$ from the 2D $\alpha$-$\beta$-BC-$\sigma_D$-$f_D$-$\Csh$ analysis. Green stars show the \emph{PSP} data, and  orange triangles the \emph{Voyager}~2 data. The horizontal black dashed line shows the value $\sigma_D \approx -1/3$ that we employ in analyses that do not constrain $\sigma_D$.}
\label{fig:sigd_contour_all}
\end{figure}

To examine the sampled posterior distributions provided by \texttt{multinest} and generate the joint-posterior and 1D posterior distributions and confidence interval plots (such as \autoref{fig:sigd_contour_all})
we make use of the public domain routines \texttt{getdist}\footnote{\url{https://github.com/cmbant/getdist}} \citep{Lewis19} and \texttt{fgivenx}\footnote{\url{https://github.com/handley-lab/fgivenx}} \citep{Handley18}.

%number of 
%number of live points as $100$ times the number of parameters we are sampling
%default sampling efficiency of 0.8
%\textcolor{red}{note: multinest expects the log-likelihood function $\ln{\mathcal{L}(\Theta)}$}
%\textcolor{red}{nested sampling}
%how it samples points (the circle stuff)
%unit hypercube that gets remapped to the prior distributions
%link to my codes

\begin{table}
\centering
\begin{tabular}{c c c}
\hline
\hline
Model & Time (minutes) & Num.~evaluations\\
\hline
\hline
%$\alpha$-$\beta$-$T_{\init}$ & 3m & 4927\\
%$\alpha$-$\beta$-initial & 9m & 21141\\
%%\red{$\alpha$-$\beta$-$T_{\init}$-$\sigma_D$-driving} & \red{NUM} & \red{NUM}\\
%$\alpha$-$\beta$-initial-$\sigma_D$-driving & 22m & 45851\\
%\emph{Helios} 2/\emph{Ulysses} dataset & 14m & 20121\\
%\emph{Voyager} 2 dataset & 23m & 38341\\
%Combined dataset & 24m & 42476\\
%No $\lambda$ $\dot{E}_{PI}$ & 24m & 60148\\
%\hline
$\alpha$-$\beta$ & 2 & 3000\\
$\alpha$-$\beta$-BC & 10 & 25000 \\
$\alpha$-$\beta$-$\sigma_D$-$f_D$-$\Csh$ & 10 & 25000 \\
$\alpha$-$\beta$-BC-$\sigma_D$-$f_D$-$\Csh$ & 40 & 90000\\
\hline
\end{tabular}
\caption{Execution times and number of likelihood evaluations for Bayesian analysis using \texttt{multinest} on a desktop computer with 32\,GB RAM, and an 8 core, 4\,GHz CPU.}
\label{mcmc:tab:time}
\end{table}

  %\input{appendices/app-tables}

%-----------------------------------
 \bibliographystyle{aasjournal}

 \bibliography{msc,
  proposal,
  bayes,
  NS_publications,
  sean_ag,
  sean_hl,
  sean_mp,
  sean_qz,
  my_lib}{}

\newcommand{\BIBand} {and} %...... how 'and' appears in authors \newcommand{\boldVol}[1] {\textbf{#1}} %...................... \providecommand{\SortNoop}[1]{} %.......Use as {\SortNoop{Aaa}} \providecommand{\sortnoop}[1]{} %.............................. \newcommand{\stereo} {\emph{{S}{T}{E}{R}{E}{O}}} %................ \providecommand{\alfven} {{A}lfv{\'e}n\ } %....................... \providecommand{\alfvenic} {{A}lfv{\'e}nic\ } %..................... \providecommand{\Alfven} {{A}lfv{\'e}n\ } %....................... \providecommand{\Alfvenic} {{A}lfv{\'e}nic\ }
\begin{thebibliography}{}
\expandafter\ifx\csname natexlab\endcsname\relax\def\natexlab#1{#1}\fi
\providecommand{\url}[1]{\href{#1}{#1}}
\providecommand{\dodoi}[1]{doi:~\href{http://doi.org/#1}{\nolinkurl{#1}}}
\providecommand{\doeprint}[1]{\href{http://ascl.net/#1}{\nolinkurl{http://ascl.net/#1}}}
\providecommand{\doarXiv}[1]{\href{https://arxiv.org/abs/#1}{\nolinkurl{https://arxiv.org/abs/#1}}}

\bibitem[{Abbott {et~al.}(2019)Abbott, Allam, Andersen, Angus, Asorey, Avelino, Avila, Bassett, Bechtol, Bernstein, Bertin, Brooks, Brout, Brown, Burke, Calcino, Carnero~Rosell, Carollo, Carrasco~Kind, Carretero, Casas, Castander, Cawthon, Challis, Childress, Clocchiatti, Cunha, D'Andrea, {da Costa}, Davis, Davis, De~Vicente, DePoy, Desai, Diehl, Doel, {Drlica-Wagner}, Eifler, Evrard, Fernandez, Filippenko, Finley, Flaugher, Foley, Fosalba, Frieman, Galbany, {Garc{\'i}a-Bellido}, Gaztanaga, Giannantonio, Glazebrook, Goldstein, {Gonz{\'a}lez-Gait{\'a}n}, Gruen, Gruendl, Gschwend, Gupta, Gutierrez, Hartley, Hinton, Hollowood, Honscheid, Hoormann, Hoyle, James, Jeltema, Johnson, Johnson, Kasai, Kent, Kessler, Kim, Kirshner, Kovacs, Krause, Kron, Kuehn, Kuhlmann, Kuropatkin, Lahav, Lasker, Lewis, Li, Lidman, Lima, Lin, Macaulay, Maia, Mandel, March, Marriner, Marshall, Martini, Menanteau, Miller, Miquel, Miranda, Mohr, Morganson, Muthukrishna, M{\"o}ller, Neilsen, Nichol, Nord, Nugent, Ogando, Palmese, Pan,
  Plazas, Pursiainen, Romer, Roodman, Rozo, Rykoff, Sako, Sanchez, Scarpine, Schindler, Schubnell, Scolnic, Serrano, {Sevilla-Noarbe}, Sharp, Smith, {Soares-Santos}, Sobreira, Sommer, Spinka, Suchyta, Sullivan, Swann, Tarle, Thomas, Thomas, Troxel, Tucker, Uddin, Walker, Wester, Wiseman, Wolf, Yanny, Zhang, Zhang, \& {DES Collaboration}}]{Abbott.etal19}
Abbott, T. M.~C., Allam, S., Andersen, P., {et~al.} 2019, The Astrophysical Journal, 872, L30, \dodoi{10.3847/2041-8213/ab04fa}

\bibitem[{Adhikari {et~al.}(2015)Adhikari, Zank, Bruno, Telloni, Hunana, Dosch, Marino, \& Hu}]{AdhikariEA15}
Adhikari, L., Zank, G.~P., Bruno, R., {et~al.} 2015, Astrophys.\ J., 805, 63, \dodoi{10.1088/0004-637X/805/1/63}

\bibitem[{Adhikari {et~al.}(2017{\natexlab{a}})Adhikari, Zank, Hunana, Shiota, Bruno, Hu, \& Telloni}]{AdhikariEA17-NI-2}
Adhikari, L., Zank, G.~P., Hunana, P., {et~al.} 2017{\natexlab{a}}, Astrophys.\ J., 841, 85, \dodoi{10.3847/1538-4357/aa6f5d}

\bibitem[{Adhikari {et~al.}(2017{\natexlab{b}})Adhikari, Zank, Telloni, Hunana, Bruno, \& Shiota}]{AdhikariEA17-NI-3}
Adhikari, L., Zank, G.~P., Telloni, D., {et~al.} 2017{\natexlab{b}}, Astrophys.\ J., 851, 117, \dodoi{10.3847/1538-4357/aa9ce4}

\bibitem[{Adhikari {et~al.}(2021)Adhikari, Zank, \& Zhao}]{AdhikariEA21-xport}
Adhikari, L., Zank, G.~P., \& Zhao, L. 2021, Fluids, 6, \dodoi{10.3390/fluids6100368}

\bibitem[{Adhikari {et~al.}(2023)Adhikari, Zank, Wang, Zhao, Telloni, Pitna, Opher, Shrestha, Mc{Comas}, \& Nykyri}]{AdhikariEA23-highBetaXport}
Adhikari, L., Zank, G.~P., Wang, B., {et~al.} 2023, Astrophys.\ J., 953, 44, \dodoi{10.3847/1538-4357/acde57}

\bibitem[{Ashton {et~al.}(2022)Ashton, Bernstein, Buchner, Chen, Cs{\'a}nyi, Fowlie, Feroz, Griffiths, Handley, Habeck, Higson, Hobson, Lasenby, Parkinson, P{\'a}rtay, Pitkin, Schneider, Speagle, South, Veitch, Wacker, Wales, \& Yallup}]{Ashton.etal22}
Ashton, G., Bernstein, N., Buchner, J., {et~al.} 2022, Nature Reviews Methods Primers, 2, 39, \dodoi{10.1038/s43586-022-00121-x}

\bibitem[{Bandyopadhyay {et~al.}(2019)Bandyopadhyay, Matthaeus, Oughton, \& Wan}]{BandyopadhyayEA19-lengths}
Bandyopadhyay, R., Matthaeus, W.~H., Oughton, S., \& Wan, M. 2019, J.\ Fluid Mech., 876, 5, \dodoi{10.1017/jfm.2019.513}

\bibitem[{Bandyopadhyay {et~al.}(2018)Bandyopadhyay, Oughton, Wan, Matthaeus, Chhiber, \& Parashar}]{BandyopadhyayEA18-Ceps}
Bandyopadhyay, R., Oughton, S., Wan, M., {et~al.} 2018, Phys.\ Rev.\ X, 8, 041052, \dodoi{10.1103/PhysRevX.8.041052}

\bibitem[{Bandyopadhyay {et~al.}(2023)Bandyopadhyay, Meyer, Matthaeus, McComas, Cranmer, Halekas, Huang, Larson, Livi, Rahmati, Whittlesey, Stevens, Kasper, \& Bale}]{BandyopadhyayEA23-peHeat}
Bandyopadhyay, R., Meyer, C.~M., Matthaeus, W.~H., {et~al.} 2023, Astrophys.\ J. Lett., 955, L28, \dodoi{10.3847/2041-8213/acf85e}

\bibitem[{Barbary(2021)}]{Barbary21}
Barbary, K. 2021, Astrophysics Source Code Library, ascl:2103.022

\bibitem[{Belcher \& Davis~Jr.(1971)}]{BelcherDavis71}
Belcher, J.~W., \& Davis~Jr., L. 1971, J.\ Geophys.\ Res., 76, 3534, \dodoi{10.1029/JA076i016p03534}

\bibitem[{Bieber {et~al.}(1996)Bieber, Wanner, \& Matthaeus}]{BieberEA96}
Bieber, J.~W., Wanner, W., \& Matthaeus, W.~H. 1996, J.\ Geophys.\ Res., 101, 2511, \dodoi{10.1029/95JA02588}

\bibitem[{Breech {et~al.}(2009)Breech, Matthaeus, Cranmer, Kasper, \& Oughton}]{BreechEA09}
Breech, B., Matthaeus, W.~H., Cranmer, S.~R., Kasper, J.~C., \& Oughton, S. 2009, J.\ Geophys.\ Res., 114, \dodoi{10.1029/2009JA014354}

\bibitem[{Breech {et~al.}(2008)Breech, Matthaeus, Minnie, Bieber, Oughton, Smith, \& Isenberg}]{BreechEA08}
Breech, B., Matthaeus, W.~H., Minnie, J., {et~al.} 2008, J.\ Geophys.\ Res., 113, \dodoi{10.1029/2007JA012711}

\bibitem[{Breech {et~al.}(2005)Breech, Matthaeus, Minnie, Oughton, Parhi, Bieber, \& Bavassano}]{BreechEA05}
---. 2005, Geophys.\ Res.\ Lett., 32, L06103, doi:10.1029/2004GL022321, \dodoi{10.1029/2004GL022321}

\bibitem[{Brewer {et~al.}(2010)Brewer, P{\'a}rtay, \& Cs{\'a}nyi}]{Brewer.etal10}
Brewer, B.~J., P{\'a}rtay, L.~B., \& Cs{\'a}nyi, G. 2010, Astrophysics Source Code Library, ascl:1010.029

\bibitem[{Bruno \& Carbone(2013)}]{BrunoCarbone13}
Bruno, R., \& Carbone, V. 2013, Living Rev.\ Solar Phys., 10, \dodoi{10.12942/lrsp-2013-2}

\bibitem[{Buchner(2016)}]{Buchner16}
Buchner, J. 2016, Astrophysics Source Code Library, ascl:1606.005

\bibitem[{Buchner(2021)}]{Buchner21}
---. 2021, The Journal of Open Source Software, 6, 3001, \dodoi{10.21105/joss.03001}

\bibitem[{Burlaga(1974)}]{Burlaga74}
Burlaga, L.~F. 1974, J.\ Geophys.\ Res., 79, 3717, \dodoi{10.1029/JA079i025p03717}

\bibitem[{Carbone \& Veltri(1990)}]{CarboneVeltri90}
Carbone, V., \& Veltri, P. 1990, Geophys.\ Astrophys.\ Fluid Dyn., 52, 153, \dodoi{10.1080/03091929008219845}

\bibitem[{Corsaro \& De~Ridder(2014)}]{Corsaro.DeRidder14}
Corsaro, E., \& De~Ridder, J. 2014, Astronomy and Astrophysics, 571, A71, \dodoi{10.1051/0004-6361/201424181}

\bibitem[{Cranmer(2009)}]{Cranmer09}
Cranmer, S.~R. 2009, Living Rev.\ Solar Phys., 6.
\newblock \url{http://www.livingreviews.org/lrsp-2009-3}

\bibitem[{Cranmer {et~al.}(2009)Cranmer, Matthaeus, Breech, \& Kasper}]{CranmerEA09}
Cranmer, S.~R., Matthaeus, W.~H., Breech, B.~A., \& Kasper, J.~C. 2009, Astrophys.\ J., 702, 1604, \dodoi{10.1088/0004-637X/702/2/1604}

\bibitem[{Cuesta {et~al.}(2022{\natexlab{a}})Cuesta, Parashar, Chhiber, \& Matthaeus}]{CuestaEA22-intermit}
Cuesta, M.~E., Parashar, T.~N., Chhiber, R., \& Matthaeus, W.~H. 2022{\natexlab{a}}, Astrophys.\ J.\ Suppl. Ser, 259, 23, \dodoi{10.3847/1538-4365/ac45fa}

\bibitem[{Cuesta {et~al.}(2022{\natexlab{b}})Cuesta, Chhiber, Roy, Goodwill, Pecora, Jarosik, Matthaeus, Parashar, \& Bandyopadhyay}]{CuestaEA22-ec}
Cuesta, M.~E., Chhiber, R., Roy, S., {et~al.} 2022{\natexlab{b}}, Astrophys.\ J. Lett., 932, L11, \dodoi{10.3847/2041-8213/ac73fd}

\bibitem[{Dryden(1943)}]{Dryden43}
Dryden, H.~L. 1943, Quart.\ Appl.\ Math., 1, 7, \dodoi{10.1090/qam/8209}

\bibitem[{Elliott {et~al.}(2019)Elliott, McComas, Zirnstein, Randol, Delamere, Livadiotis, Bagenal, Barnes, Stern, Young, {et~al.}}]{ElliottEA19}
Elliott, H.~A., McComas, D.~J., Zirnstein, E.~J., {et~al.} 2019, The Astrophysical Journal, 885, 156, \dodoi{10.3847/1538-4357/ab3e49}

\bibitem[{Ellison(2004)}]{Ellison04}
Ellison, A.~M. 2004, Ecology Letters, 7, 509, \dodoi{https://doi.org/10.1111/j.1461-0248.2004.00603.x}

\bibitem[{Engelbrecht {et~al.}(2022)Engelbrecht, Effenberger, Florinski, Potgieter, Ruffolo, Chhiber, Usmanov, Rankin, \& Els}]{Engelbrecht.etal22}
Engelbrecht, N.~E., Effenberger, F., Florinski, V., {et~al.} 2022, Space Sci Rev, 218, 33, \dodoi{10.1007/s11214-022-00896-1}

\bibitem[{Feroz \& Hobson(2008)}]{Feroz.Hobson08}
Feroz, F., \& Hobson, M.~P. 2008, Monthly Notices of the Royal Astronomical Society, 384, 449, \dodoi{10.1111/j.1365-2966.2007.12353.x}

\bibitem[{Feroz {et~al.}(2009)Feroz, Hobson, \& Bridges}]{Feroz.etal09}
Feroz, F., Hobson, M.~P., \& Bridges, M. 2009, Monthly Notices of the Royal Astronomical Society, 398, 1601, \dodoi{10.1111/j.1365-2966.2009.14548.x}

\bibitem[{Fraternale {et~al.}(2022)Fraternale, Adhikari, Fichtner, Kim, Kleimann, Oughton, Pogorelov, Roytershteyn, Smith, Usmanov, Zank, \& Zhao}]{FraternaleEA22-ssr}
Fraternale, F., Adhikari, L., Fichtner, H., {et~al.} 2022, Space Sci.\ Rev., 218, 50, \dodoi{10.1007/s11214-022-00914-2}

\bibitem[{Goldreich \& Sridhar(1997)}]{GoldreichSridhar97}
Goldreich, P., \& Sridhar, S. 1997, Astrophys.\ J., 485, 680, \dodoi{10.1086/304442}

\bibitem[{Handley(2018)}]{Handley18}
Handley, W. 2018, The Journal of Open Source Software, 3, 849, \dodoi{10.21105/joss.00849}

\bibitem[{Handley {et~al.}(2015)Handley, Hobson, \& Lasenby}]{Handley.etal15}
Handley, W.~J., Hobson, M.~P., \& Lasenby, A.~N. 2015, Monthly Notices of the Royal Astronomical Society, 453, 4384, \dodoi{10.1093/mnras/stv1911}

\bibitem[{Hellinger {et~al.}(2013)Hellinger, {Tr{\'a}vn\'{\i}{\v c}ek}, {{\v{S}}tver{\'a}k}, Matteini, \& Velli}]{HellingerEA13}
Hellinger, P., {Tr{\'a}vn\'{\i}{\v c}ek}, P.~M., {{\v{S}}tver{\'a}k}, v., Matteini, L., \& Velli, M. 2013, J.\ Geophys.\ Res., 118, 1351, \dodoi{10.1002/jgra.50107}

\bibitem[{Hojjati {et~al.}(2011)Hojjati, Pogosian, \& Zhao}]{Hojjati.etal11}
Hojjati, A., Pogosian, L., \& Zhao, G.-B. 2011, Journal of Cosmology and Astroparticle Physics, 2011, 005, \dodoi{10.1088/1475-7516/2011/08/005}

\bibitem[{Hossain {et~al.}(1995)Hossain, Gray, Pontius~Jr., Matthaeus, \& Oughton}]{HossainEA95}
Hossain, M., Gray, P.~C., Pontius~Jr., D.~H., Matthaeus, W.~H., \& Oughton, S. 1995, Phys.\ Fluids, 7, 2886, \dodoi{10.1063/1.868665}

\bibitem[{Howes(2011)}]{Howes11}
Howes, G.~G. 2011, Astrophys.\ J., 738, 40, \dodoi{10.1088/0004-637X/738/1/40}

\bibitem[{Huang {et~al.}(2019)Huang, Shao, Wu, Beck, \& Li}]{Huang.etal19}
Huang, Y., Shao, C., Wu, B., Beck, J.~L., \& Li, H. 2019, Advances in Structural Engineering, 22, 1329, \dodoi{10.1177/1369433218811540}

\bibitem[{Isenberg(2005)}]{Isenberg05}
Isenberg, P.~A. 2005, Astrophys.\ J., 623, 502, \dodoi{10.1086/428609}

\bibitem[{Isenberg {et~al.}(2003)Isenberg, Smith, \& Matthaeus}]{IsenbergEA03}
Isenberg, P.~A., Smith, C.~W., \& Matthaeus, W.~H. 2003, Astrophys.\ J., 592, 564, \dodoi{10.1086/375584}

\bibitem[{Isenberg {et~al.}(2010)Isenberg, Smith, Matthaeus, \& Richardson}]{IsenbergEA10}
Isenberg, P.~A., Smith, C.~W., Matthaeus, W.~H., \& Richardson, J.~D. 2010, Astrophys.\ J., 719, 716, \dodoi{10.1088/0004-637X/719/1/716}

\bibitem[{Javid {et~al.}(2019)Javid, Perrott, Rumsey, \& Saunders}]{Javid.etal19}
Javid, K., Perrott, Y.~C., Rumsey, C., \& Saunders, R. D.~E. 2019, Monthly Notices of the Royal Astronomical Society, 489, 3135, \dodoi{10.1093/mnras/stz2341}

\bibitem[{Jeffreys \& Jeffreys(1998)}]{Jeffreys.Jeffreys98}
Jeffreys, S.~H., \& Jeffreys, S.~H. 1998, The {{Theory}} of {{Probability}}, third edition, third edition edn., Oxford {{Classic Texts}} in the {{Physical Sciences}} ({Oxford, New York}: {Oxford University Press})

\bibitem[{Jungman {et~al.}(1996)Jungman, Kamionkowski, Kosowsky, \& Spergel}]{Jungman.etal96}
Jungman, G., Kamionkowski, M., Kosowsky, A., \& Spergel, D.~N. 1996, Physical Review D, 54, 1332, \dodoi{10.1103/PhysRevD.54.1332}

\bibitem[{Kester \& Mueller(2021)}]{Kester.Mueller21}
Kester, D., \& Mueller, M. 2021, {{BayesicFitting}}, a {{PYTHON Toolbox}} for {{Bayesian Fitting}} and {{Evidence Calculation}}, \dodoi{10.48550/arXiv.2109.11976}

\bibitem[{Kleimann {et~al.}(2023)Kleimann, Oughton, Fichtner, \& Scherer}]{KleimannEA23}
Kleimann, J., Oughton, S., Fichtner, H., \& Scherer, K. 2023, Astrophys.\ J., 953, 133, \dodoi{10.3847/1538-4357/acd84e}

\bibitem[{Kolmogorov(1941)}]{Kol41b}
Kolmogorov, A.~N. 1941, C.R. Acad. Sci. U.R.S.S., 31, 538

\bibitem[{Lee \& Ip(1987)}]{LeeIp87}
Lee, M., \& Ip, W.-H. 1987, J.\ Geophys.\ Res., 92, 11041, \dodoi{10.1029/JA092iA10p11041}

\bibitem[{Lewis(2019)}]{Lewis19}
Lewis, A. 2019, {{GetDist}}: A {{Python}} Package for Analysing {{Monte Carlo}} Samples, \dodoi{10.48550/arXiv.1910.13970}

\bibitem[{Linkmann {et~al.}(2017)Linkmann, Berera, \& Goldstraw}]{LinkmannEA17-Ceps}
Linkmann, M., Berera, A., \& Goldstraw, E.~E. 2017, Phys.\ Rev.\ E, 95, 013102, \dodoi{10.1103/PhysRevE.95.013102}

\bibitem[{Matthaeus {et~al.}(1996{\natexlab{a}})Matthaeus, Ghosh, Oughton, \& Roberts}]{MattEA96-var}
Matthaeus, W.~H., Ghosh, S., Oughton, S., \& Roberts, D.~A. 1996{\natexlab{a}}, J.\ Geophys.\ Res., 101, 7619, \dodoi{10.1029/95JA03830}

\bibitem[{Matthaeus {et~al.}(1990)Matthaeus, Goldstein, \& Roberts}]{MattEA90}
Matthaeus, W.~H., Goldstein, M.~L., \& Roberts, D.~A. 1990, J.\ Geophys.\ Res., 95, 20\,673, \dodoi{10.1029/JA095iA12p20673}

\bibitem[{Matthaeus {et~al.}(1994)Matthaeus, Oughton, Pontius, \& Zhou}]{MattEA94-ec}
Matthaeus, W.~H., Oughton, S., Pontius, D., \& Zhou, Y. 1994, J.\ Geophys.\ Res., 99, 19\,267, \dodoi{10.1029/94JA01233}

\bibitem[{Matthaeus {et~al.}(1996{\natexlab{b}})Matthaeus, Zank, \& Oughton}]{MattEA96-jpp}
Matthaeus, W.~H., Zank, G.~P., \& Oughton, S. 1996{\natexlab{b}}, J.\ Plasma Phys., 56, 659, \dodoi{10.1017/S0022377800019516}

\bibitem[{Matthaeus {et~al.}(1999)Matthaeus, Zank, Smith, \& Oughton}]{MattEA99-swh}
Matthaeus, W.~H., Zank, G.~P., Smith, C.~W., \& Oughton, S. 1999, Phys.\ Rev.\ Lett., 82, 3444, \dodoi{10.1103/PhysRevLett.82.3444}

\bibitem[{McComas {et~al.}(2019)McComas, Rankin, Schwadron, \& Swaczyna}]{McComasEA19}
McComas, D., Rankin, J., Schwadron, N., \& Swaczyna, P. 2019, The Astrophysical Journal, 884, 145

\bibitem[{Neugebauer {et~al.}(1995)Neugebauer, Goldstein, Mc{Comas}, Suess, \& Balogh}]{NeugebauerEA95}
Neugebauer, M., Goldstein, B.~E., Mc{Comas}, D.~J., Suess, S.~T., \& Balogh, A. 1995, J.\ Geophys.\ Res., 100, 23\,389, \dodoi{10.1029/95JA02723}

\bibitem[{Oughton \& Bishop(2024)}]{OughtonBishop-sw16}
Oughton, S., \& Bishop, M.~A. 2024, in Solar Wind 16, Vol. in preparation

\bibitem[{Oughton \& Engelbrecht(2021)}]{OughtonEngelbrecht21}
Oughton, S., \& Engelbrecht, N.~E. 2021, New Astron., 83, \dodoi{10.1016/j.newast.2020.101507}

\bibitem[{Oughton \& Matthaeus(2020)}]{OughtonMatt20}
Oughton, S., \& Matthaeus, W.~H. 2020, Astrophys.\ J., 897, \dodoi{10.3847/1538-4357/ab8f2a}

\bibitem[{Oughton {et~al.}(2011)Oughton, Matthaeus, Smith, Breech, \& Isenberg}]{OughtonEA11}
Oughton, S., Matthaeus, W.~H., Smith, C.~W., Breech, B., \& Isenberg, P.~A. 2011, J.\ Geophys.\ Res., 116, \dodoi{10.1029/2010JA016365}

\bibitem[{Oughton {et~al.}(2015)Oughton, Matthaeus, Wan, \& Osman}]{OughtonEA15}
Oughton, S., Matthaeus, W.~H., Wan, M., \& Osman, K.~T. 2015, Phil.\ Trans.\ R.\ Soc.\ A, 373, \dodoi{10.1098/rsta.2014.0152}

\bibitem[{Oughton {et~al.}(1994)Oughton, Priest, \& Matthaeus}]{OughtonEA94}
Oughton, S., Priest, E.~R., \& Matthaeus, W.~H. 1994, J.\ Fluid Mech., 280, 95, \dodoi{10.1017/S0022112094002867}

\bibitem[{Parashar {et~al.}(2019)Parashar, Cuesta, \& Matthaeus}]{ParasharEA19}
Parashar, T.~N., Cuesta, M., \& Matthaeus, W.~H. 2019, Astrophys.\ J., 884, L57, \dodoi{10.3847/2041-8213/ab4a82}

\bibitem[{Parker(1958)}]{Parker58}
Parker, E.~N. 1958, Astrophys.\ J., 128, 664, \dodoi{10.1086/146579}

\bibitem[{Parker(1965)}]{Parker65-wkb}
---. 1965, Space Sci.\ Rev., 4, 666, \dodoi{10.1007/BF00216273}

\bibitem[{Parkinson \& Liddle(2013)}]{Parkinson.Liddle13}
Parkinson, D., \& Liddle, A.~R. 2013, Statistical Analysis and Data Mining: The ASA Data Science Journal, 9, 3, \dodoi{10.1002/sam.11179}

\bibitem[{Parkinson {et~al.}(2011)Parkinson, Mukherjee, \& Liddle}]{Parkinson.etal11}
Parkinson, D., Mukherjee, P., \& Liddle, A. 2011, Astrophysics Source Code Library, ascl:1110.019

\bibitem[{Pei {et~al.}(2010)Pei, Bieber, Breech, Burger, Clem, \& Matthaeus}]{PeiEA10}
Pei, C., Bieber, J.~W., Breech, B., {et~al.} 2010, J.\ Geophys.\ Res., 115, \dodoi{10.1029/2009JA014705}

\bibitem[{Perlmutter {et~al.}(1995)Perlmutter, Pennypacker, Goldhaber, Goobar, Muller, Newberg, Desai, Kim, Kim, Small, Boyle, Crawford, McMahon, Bunclark, Carter, Irwin, Terlevich, Ellis, Glazebrook, Couch, Mould, Small, \& Abraham}]{Perlmutter.etal95}
Perlmutter, S., Pennypacker, C.~R., Goldhaber, G., {et~al.} 1995, The Astrophysical Journal, 440, L41, \dodoi{10.1086/187756}

\bibitem[{Perri \& Balogh(2010)}]{PerriBalogh10-sigc}
Perri, S., \& Balogh, A. 2010, Geophys.\ Res.\ Lett., 37, \dodoi{10.1029/2010GL044570}

\bibitem[{Perrott {et~al.}(2019)Perrott, Javid, Carvalho, Elwood, Hobson, Lasenby, Olamaie, \& Saunders}]{Perrott.etal19}
Perrott, Y.~C., Javid, K., Carvalho, P., {et~al.} 2019, Monthly Notices of the Royal Astronomical Society, 486, 2116, \dodoi{10.1093/mnras/stz826}

\bibitem[{Petzold(1983)}]{LSODE}
Petzold, L. 1983, SIAM Journal on Scientific and Statistical Computing, 4, 136, \dodoi{10.1137/0904010}

\bibitem[{{Planck Collaboration} {et~al.}(2020){Planck Collaboration}, Aghanim, Akrami, Ashdown, Aumont, Baccigalupi, Ballardini, Banday, Barreiro, Bartolo, Basak, Battye, Benabed, Bernard, Bersanelli, Bielewicz, Bock, Bond, Borrill, Bouchet, Boulanger, Bucher, Burigana, Butler, Calabrese, Cardoso, Carron, Challinor, Chiang, Chluba, Colombo, Combet, Contreras, Crill, Cuttaia, {de Bernardis}, {de Zotti}, Delabrouille, Delouis, Di~Valentino, Diego, Dor{\'e}, Douspis, Ducout, Dupac, Dusini, Efstathiou, Elsner, En{\ss}lin, Eriksen, Fantaye, Farhang, Fergusson, {Fernandez-Cobos}, Finelli, Forastieri, Frailis, Fraisse, Franceschi, Frolov, Galeotta, Galli, Ganga, {G{\'e}nova-Santos}, Gerbino, Ghosh, {Gonz{\'a}lez-Nuevo}, G{\'o}rski, Gratton, Gruppuso, Gudmundsson, Hamann, Handley, Hansen, Herranz, Hildebrandt, Hivon, Huang, Jaffe, Jones, Karakci, Keih{\"a}nen, Keskitalo, Kiiveri, Kim, Kisner, Knox, Krachmalnicoff, Kunz, {Kurki-Suonio}, Lagache, Lamarre, Lasenby, Lattanzi, Lawrence, Le~Jeune, Lemos, Lesgourgues,
  Levrier, Lewis, Liguori, Lilje, Lilley, Lindholm, {L{\'o}pez-Caniego}, Lubin, Ma, {Mac{\'i}as-P{\'e}rez}, Maggio, Maino, Mandolesi, Mangilli, {Marcos-Caballero}, Maris, Martin, Martinelli, {Mart{\'i}nez-Gonz{\'a}lez}, Matarrese, Mauri, McEwen, Meinhold, Melchiorri, Mennella, Migliaccio, Millea, Mitra, {Miville-Desch{\^e}nes}, Molinari, Montier, Morgante, Moss, Natoli, {N{\o}rgaard-Nielsen}, Pagano, Paoletti, Partridge, Patanchon, Peiris, Perrotta, Pettorino, Piacentini, Polastri, Polenta, Puget, Rachen, Reinecke, Remazeilles, Renzi, Rocha, Rosset, Roudier, {Rubi{\~n}o-Mart{\'i}n}, {Ruiz-Granados}, Salvati, Sandri, Savelainen, Scott, Shellard, Sirignano, Sirri, Spencer, Sunyaev, {Suur-Uski}, Tauber, Tavagnacco, Tenti, Toffolatti, Tomasi, Trombetti, Valenziano, Valiviita, Van~Tent, Vibert, Vielva, Villa, Vittorio, Wandelt, Wehus, White, White, Zacchei, \& Zonca}]{PlanckCollaboration.etal20}
{Planck Collaboration}, Aghanim, N., Akrami, Y., {et~al.} 2020, Astronomy and Astrophysics, 641, A6, \dodoi{10.1051/0004-6361/201833910}

\bibitem[{Pogorelov {et~al.}(2024)Pogorelov, Arge, Caplan, Colella, Linker, Singh, Straalen, Upton, Downs, Gebhart, Hegde, Henney, Jones, Johnston, Kim, Marble, Raza, Stulajter, \& Turtle}]{Pogorelov.etal24}
Pogorelov, N.~V., Arge, C.~N., Caplan, R.~M., {et~al.} 2024, J. Phys.: Conf. Ser., 2742, 012013, \dodoi{10.1088/1742-6596/2742/1/012013}

\bibitem[{Richardson \& Smith(2003)}]{RichardsonSmith03}
Richardson, J.~D., \& Smith, C.~W. 2003, Geophys.\ Res.\ Lett., 30, 1206, doi:10.1029/2002GL016551, \dodoi{10.1029/2002GL016551}

\bibitem[{Saffman(1967)}]{Saffman67b}
Saffman, P.~G. 1967, Phys.\ Fluids, 10, 1349, \dodoi{10.1080/00107510802066753}

\bibitem[{Schekochihin(2022)}]{Schekochihin-biased}
Schekochihin, A.~A. 2022, J.\ Plasma Phys., 88, 155880501, \dodoi{10.1017/S0022377822000721}

\bibitem[{Schwenn(2006)}]{Schwenn06}
Schwenn, R. 2006, Living Rev. Sol. Phys., 3, 2, \dodoi{10.12942/lrsp-2006-2}

\bibitem[{Sharma(2017)}]{Sharma17}
Sharma, S. 2017, Annual Review of Astronomy and Astrophysics, 55, 213, \dodoi{10.1146/annurev-astro-082214-122339}

\bibitem[{Shebalin {et~al.}(1983)Shebalin, Matthaeus, \& Montgomery}]{ShebalinEA83}
Shebalin, J.~V., Matthaeus, W.~H., \& Montgomery, D. 1983, J.\ Plasma Phys., 29, 525, \dodoi{10.1017/S0022377800000933}

\bibitem[{Shiota {et~al.}(2017)Shiota, Zank, Adhikari, Hunana, Telloni, \& Bruno}]{ShiotaEA17}
Shiota, D., Zank, G.~P., Adhikari, L., {et~al.} 2017, Astrophys.\ J., 837, 75, \dodoi{10.3847/1538-4357/aa60bc}

\bibitem[{Skilling(2004)}]{Skilling04}
Skilling, J. 2004, in BAYESIAN INFERENCE AND MAXIMUM ENTROPY METHODS IN SCIENCE AND ENGINEERING: 24th International Workshop on {Bayesian} Inference and Maximum Entropy Methods in Science and Engineering, Vol. 735 (AIP), 395--405, \dodoi{10.1063/1.1835238}

\bibitem[{Smith {et~al.}(2006{\natexlab{a}})Smith, Isenberg, Matthaeus, \& Richardson}]{SmithEA06-pi}
Smith, C.~W., Isenberg, P.~A., Matthaeus, W.~H., \& Richardson, J.~D. 2006{\natexlab{a}}, Astrophys.\ J., 638, 508, \dodoi{10.1086/498671}

\bibitem[{Smith {et~al.}(2001)Smith, Matthaeus, Zank, Ness, Oughton, \& Richardson}]{SmithEA01}
Smith, C.~W., Matthaeus, W.~H., Zank, G.~P., {et~al.} 2001, J.\ Geophys.\ Res., 106, 8253, \dodoi{10.1029/2000JA000366}

\bibitem[{Smith \& Vasquez(2024)}]{SmithVasquez24}
Smith, C.~W., \& Vasquez, B.~J. 2024, Front. Astron. Space Sci., 11, \dodoi{10.3389/fspas.2024.1371058}

\bibitem[{Smith {et~al.}(2006{\natexlab{b}})Smith, Vasquez, \& Hamilton}]{SmithEA06-aniso}
Smith, C.~W., Vasquez, B.~J., \& Hamilton, K. 2006{\natexlab{b}}, J.\ Geophys.\ Res., 111, A09111, doi:10.1029/2006JA011651, \dodoi{10.1029/2006JA011651}

\bibitem[{Smith \& Wolfe(1976)}]{SmithWolfe76}
Smith, E.~J., \& Wolfe, J.~H. 1976, Geophys.\ Res.\ Lett., 3, 137, \dodoi{10.1029/GL003i003p00137}

\bibitem[{Sok{\'o}{\l} {et~al.}(2022)Sok{\'o}{\l}, Kucharek, Baliukin, Fahr, Izmodenov, Kornbleuth, Mostafavi, Opher, Park, Pogorelov, Quinn, Smith, Zank, \& Zhang}]{Sokol.etal22}
Sok{\'o}{\l}, J.~M., Kucharek, H., Baliukin, I.~I., {et~al.} 2022, Space Sci Rev, 218, 18, \dodoi{10.1007/s11214-022-00883-6}

\bibitem[{Speagle(2020)}]{Speagle20}
Speagle, J.~S. 2020, Monthly Notices of the Royal Astronomical Society, 493, 3132, \dodoi{10.1093/mnras/staa278}

\bibitem[{Sreenivasan(1998)}]{Sreenivasan98}
Sreenivasan, K.~R. 1998, Phys.\ Fluids, 10, 528, \dodoi{10.1080/00107510802066753}

\bibitem[{Trotta(2008)}]{trotta2008bayes}
Trotta, R. 2008, Contemporary Physics, 49, 71, \dodoi{10.1080/00107510802066753}

\bibitem[{Tu(1987)}]{Tu87}
Tu, C. 1987, Solar Phys., 109, 149, \dodoi{10.1086/498671}

\bibitem[{Tu \& Marsch(1995)}]{TuMarsch95}
Tu, C.-Y., \& Marsch, E. 1995, Space Sci.\ Rev., 73, 1, \dodoi{10.1007/BF00748891}

\bibitem[{Tu {et~al.}(1984)Tu, Pu, \& Wei}]{TuEA84}
Tu, C.-Y., Pu, Z.-Y., \& Wei, F.-S. 1984, J.\ Geophys.\ Res., 89, 9695, \dodoi{10.1029/JA089iA11p09695}

\bibitem[{Usmanov {et~al.}(2011)Usmanov, Matthaeus, Breech, \& Goldstein}]{UsmanovEA11}
Usmanov, A., Matthaeus, W.~H., Breech, B.~A., \& Goldstein, M.~L. 2011, Astrophys.\ J., 727, 84, \dodoi{10.1088/0004-637X/727/2/84}

\bibitem[{Usmanov {et~al.}(2014)Usmanov, Goldstein, \& Matthaeus}]{UsmanovEA14}
Usmanov, A.~V., Goldstein, M.~L., \& Matthaeus, W.~H. 2014, Astrophys.\ J., 788, 43, \dodoi{10.1088/0004-637X/788/1/43}

\bibitem[{Usmanov {et~al.}(2016)Usmanov, Goldstein, \& Matthaeus}]{UsmanovEA16}
---. 2016, Astrophys.\ J., 820, 17, \dodoi{10.3847/0004-637X/820/1/17}

\bibitem[{Usmanov {et~al.}(2018)Usmanov, Matthaeus, Goldstein, \& Chhiber}]{UsmanovEA18}
Usmanov, A.~V., Matthaeus, W.~H., Goldstein, M.~L., \& Chhiber, R. 2018, Astrophys.\ J., 865, 25, \dodoi{10.3847/1538-4357/aad687}

\bibitem[{van~de Schoot {et~al.}(2021)van~de Schoot, Depaoli, King, Kramer, M{\"a}rtens, Tadesse, Vannucci, Gelman, Veen, Willemsen, {et~al.}}]{vandeschoot.etal21}
van~de Schoot, R., Depaoli, S., King, R., {et~al.} 2021, Nature Reviews Methods Primers, 1, 1

\bibitem[{{van der}~{Holst} {et~al.}(2014){van der}~{Holst}, Sokolov, Meng, Jin, Manchester, T{\'o}th, \& Gombosi}]{vanderHolstEA14}
{van der}~{Holst}, B., Sokolov, I.~V., Meng, X., {et~al.} 2014, Astrophys.\ J., 782, 81, \dodoi{10.1088/0004-637X/782/2/81}

\bibitem[{Velli {et~al.}(1989)Velli, Grappin, \& Mangeney}]{VelliEA89}
Velli, M., Grappin, R., \& Mangeney, A. 1989, Phys.\ Rev.\ Lett., 63, 1807, \dodoi{10.1103/PhysRevLett.63.1807}

\bibitem[{von Toussaint(2011)}]{vontoussaint11}
von Toussaint, U. 2011, Rev. Mod. Phys., 83, 943, \dodoi{10.1103/RevModPhys.83.943}

\bibitem[{Wan {et~al.}(2012)Wan, Oughton, Servidio, \& Matthaeus}]{WanEA12-vKH}
Wan, M., Oughton, S., Servidio, S., \& Matthaeus, W.~H. 2012, J.\ Fluid Mech., 697, 296, \dodoi{10.1017/jfm.2012.61}

\bibitem[{Wang \& Richardson(2003)}]{WangRichardson03}
Wang, C., \& Richardson, J. 2003, Journal of Geophysical Research: Space Physics, 108, \dodoi{https://doi.org/10.1029/2002JA009322}

\bibitem[{Wiengarten {et~al.}(2015)Wiengarten, Fichtner, Kleimann, \& Kissmann}]{WiengartenEA15}
Wiengarten, T., Fichtner, H., Kleimann, J., \& Kissmann, R. 2015, Astrophys.\ J., 805, 155, \dodoi{10.1088/0004-637X/805/2/155}

\bibitem[{Wiengarten {et~al.}(2016)Wiengarten, Oughton, Engelbrecht, Fichtner, Kleimann, \& Scherer}]{WiengartenEA16}
Wiengarten, T., Oughton, S., Engelbrecht, N.~E., {et~al.} 2016, Astrophys.\ J., 833, 17, \dodoi{10.3847/0004-637X/833/1/17}

\bibitem[{Williams \& Zank(1994)}]{WilliamsZank94}
Williams, L.~L., \& Zank, G.~P. 1994, J.\ Geophys.\ Res., 99, 19\,229, \dodoi{10.1029/94JA01657}

\bibitem[{Wrench {et~al.}(2024)Wrench, Parashar, Oughton, {de}~Lange, \& Frean}]{WrenchEA24}
Wrench, D., Parashar, T.~N., Oughton, S., {de}~Lange, K., \& Frean, M. 2024, Astrophys.\ J., 961, 182, \dodoi{10.3847/1538-4357/ad118e}

\bibitem[{Yokoi {et~al.}(2008)Yokoi, Rubinstein, Hamba, \& Yoshizawa}]{YokoiEA08}
Yokoi, N., Rubinstein, R., Hamba, F., \& Yoshizawa, A. 2008, J.\ Turb., 9, 1, \dodoi{10.1080/14685240802433057}

\bibitem[{Zank {et~al.}(2017)Zank, Adhikari, Hunana, Shiota, Bruno, \& Telloni}]{ZankEA17-NIxport}
Zank, G.~P., Adhikari, L., Hunana, P., {et~al.} 2017, Astrophys.\ J., 835, 147, \dodoi{10.3847/1538-4357/835/2/147}

\bibitem[{Zank {et~al.}(2018)Zank, Adhikari, Zhao, Mostafavi, Zirnstein, \& Mc{Comas}}]{ZankEA18-pui}
Zank, G.~P., Adhikari, L., Zhao, L.-L., {et~al.} 2018, Astrophys.\ J., 869, 23, \dodoi{10.3847/1538-4357/aaebfe}

\bibitem[{Zank {et~al.}(2012)Zank, Dosch, Hunana, Florinski, Matthaeus, \& Webb}]{ZankEA12-xport}
Zank, G.~P., Dosch, A., Hunana, P., {et~al.} 2012, Astrophys.\ J., 745, 35, \dodoi{10.1088/0004-637X/745/1/35}

\bibitem[{Zank {et~al.}(1996)Zank, Matthaeus, \& Smith}]{ZankEA96}
Zank, G.~P., Matthaeus, W.~H., \& Smith, C.~W. 1996, J.\ Geophys.\ Res., 101, 17\,093, \dodoi{10.1029/96JA01275}

\bibitem[{Zhou \& Matthaeus(1990)}]{ZhouMatt90a}
Zhou, Y., \& Matthaeus, W.~H. 1990, J.\ Geophys.\ Res., 95, 10\,291, \dodoi{10.1029/JA095iA07p10291}

\bibitem[{Zirnstein {et~al.}(2022)Zirnstein, M{\"o}bius, Zhang, Bower, Elliott, McComas, Pogorelov, \& Swaczyna}]{Zirnstein.etal22}
Zirnstein, E.~J., M{\"o}bius, E., Zhang, M., {et~al.} 2022, Space Sci Rev, 218, 28, \dodoi{10.1007/s11214-022-00895-2}

\end{thebibliography}

\end{document}